\documentclass[aps,amsmath,amssymb,preprint,showkeys,superscriptaddress,showpacs,floatfix,longbibliography]{revtex4-1}
\usepackage{graphicx}
\usepackage{dcolumn}
\usepackage[colorlinks=true,linkcolor=blue ,citecolor=blue,urlcolor=blue]{hyperref}
\usepackage{multirow}
\usepackage[usenames,dvipsnames]{xcolor}
\usepackage{soul}
\usepackage{braket}
\usepackage{bm}
\usepackage{nicefrac}
\usepackage{siunitx}
\usepackage{color}
\usepackage[utf8]{inputenc}
\usepackage{upgreek}

\newcommand{\HH}{\mathcal{H}}

\newcommand{\iu}{\mathrm{i}}

\newcommand{\VEC}[1]{\mathbf{#1}}

\newcommand{\dv}[2]{\frac{\textnormal{d}#1}{\textnormal{d}#2}}
\newcommand{\partdv}[2]{\frac{\partial#1}{\partial#2}}

\newcommand{\sgn}{\text{sgn}} 
\DeclareSIUnit\ML{ML}
\DeclareSIUnit\MLs{MLs}
\DeclareSIUnit\meVA{meV\angstrom^2}

\begin{document}

\title{Nonreciprocity of spin waves in noncollinear magnets due to the Dzyaloshinskii-Moriya interaction}

\author{Flaviano José dos Santos}\email{f.dos.santos@fz-juelich.de}
\affiliation{Peter Gr\"{u}nberg Institut and Institute for Advanced Simulation, Forschungszentrum J\"{u}lich and JARA, D-52425 J\"{u}lich, Germany}
\affiliation{Department of Physics, RWTH Aachen University, 52056 Aachen, Germany}
\author{Manuel dos Santos Dias}
\author{Samir Lounis}
\affiliation{Peter Gr\"{u}nberg Institut and Institute for Advanced Simulation, Forschungszentrum J\"{u}lich and JARA, D-52425 J\"{u}lich, Germany}

\date{\today}

\begin{abstract}

Broken inversion symmetry in combination with the spin-orbit interaction generates a finite Dzyaloshinskii-Moriya interaction (DMI), which can induce noncollinear spin textures of chiral nature.
The DMI is characterized by an interaction vector whose magnitude, direction and symmetries are crucial to determine the stability of various spin textures, such as skyrmions and spin spirals.
The DMI can be measured from the nonreciprocity of spin waves in ferromagnets, which can be probed via inelastic scattering experiments.
In a ferromagnet, the DMI can modify the spin-wave dispersion, moving its minimum away from the $\Gamma$ point.
Spin waves propagating with opposite wavevectors are then characterized by different group velocities, energies and lifetimes, defining their nonreciprocity.
Here, we address the case of complex spin textures, where the manifestation of  DMI-induced chiral asymmetries remains to be explored.
We discuss such nonreciprocal effects and propose ways of accessing the magnitude and direction of the DMI vectors in the context of spin-polarized or spin-resolved inelastic scattering experiments.
We show that only when a periodic magnetic system has finite net magnetization, that is, when the vector sum of all magnetic moments is nonzero, can it present a total nonreciprocal spin-wave spectrum.
However, even zero-net-magnetization systems, such as collinear antiferromagnets and cycloidal spin spirals, can have spin-wave modes that are individually nonreciprocal, while the total spectrum remains reciprocal.

\end{abstract}

\maketitle

\section{Introduction}

In the publication ``More is different" from 1972, P. W. Anderson discusses the importance of symmetry breaking in nature~\cite{anderson_more_1972}. Since then, there has been an ever-growing interest in the symmetries and symmetry-breaking of condensed-matter systems.
An example is the Dzyaloshinskii-Moriya interaction (DMI), which originates from the combination of broken inversion symmetry with the spin-orbit interaction~\cite{dzyaloshinsky_thermodynamic_1958,moriya_anisotropic_1960}.
The DMI is a chiral interaction introducing a vector coupling between two spin moments, $\VEC{D}_{12} \cdot (\VEC{S}_1 \times \VEC{S}_2)$, which favors one sense of rotation of the spins.
Thus, some static and dynamical physical properties of magnetic materials can acquire the chirality of the DMI.
For example, spin-polarized scanning tunneling microscopy revealed that spin spirals with a unique rotational sense are present in a single atomic layer of manganese deposited on tungsten~\cite{bode_chiral_2007,ferriani_atomic-scale_2008}. 
Also, when a spin-wave current is driven by a thermal gradient, the DMI can lead to the magnon Hall effect~\cite{onose_observation_2010}.
In ferromagnetic materials, the DMI can impart a fixed chirality to the domain walls, which can then be moved very efficiently with applied currents~\cite{ryu_chiral_2013,emori_current-driven_2013}.
Moreover, the DMI is often the stabilizing mechanism for magnetic skyrmions, which are noncollinear spin textures with particle-like properties currently under heavy investigation as potential future bits for data storage devices~\cite{fert_skyrmions_2013,sampaio_nucleation_2013,tomasello_strategy_2014,zhou_reversible_2014,crum_perpendicular_2015,zhang_magnetic_2015,yu_room-temperature_2016,garcia-sanchez_skyrmion-based_2016,xia_control_2017,garst_collective_2017,adams_response_2018,rozsa_effective_2018,zazvorka_thermal_2019,diaz_topological_2019}.
Whether a skyrmion or an antiskyrmion (spin textures with the same polarity but opposite vorticity~\cite{nagaosa_topological_2013}) can be stabilized is determined by the chirality and symmetries of the DMI~\cite{bogdanov_thermodynamically_1989,rosler_spontaneous_2006,hoffmann_antiskyrmions_2017}.
Thus, the knowledge of the DMI is essential to understand, design and control many properties of magnetic systems.
However, the DMI itself cannot be directly measured.
Instead, we observe DMI-dependent properties, which in turn allow us to obtain information about the DMI for a particular system.
Therefore, it is crucial to discover better and more complete ways to experimentally characterize this interaction in complex magnetic materials, as a way of exploiting chirality-dependent effects~\cite{cheong_broken_2018}.

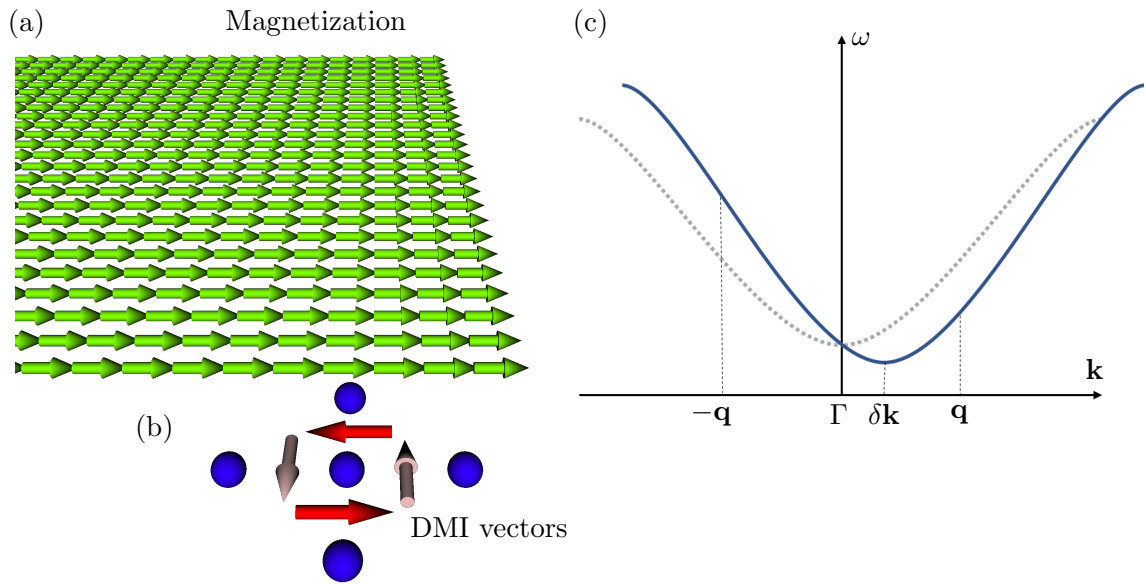
\begin{figure}[t]
\setlength{\unitlength}{1cm} 
\newcommand{\boxsize}{0.3}

  \begin{picture}(15,8.5)
    \put(0.0, 0){ \includegraphics[width=7.2 cm,trim={0 0 500 300},clip=true]{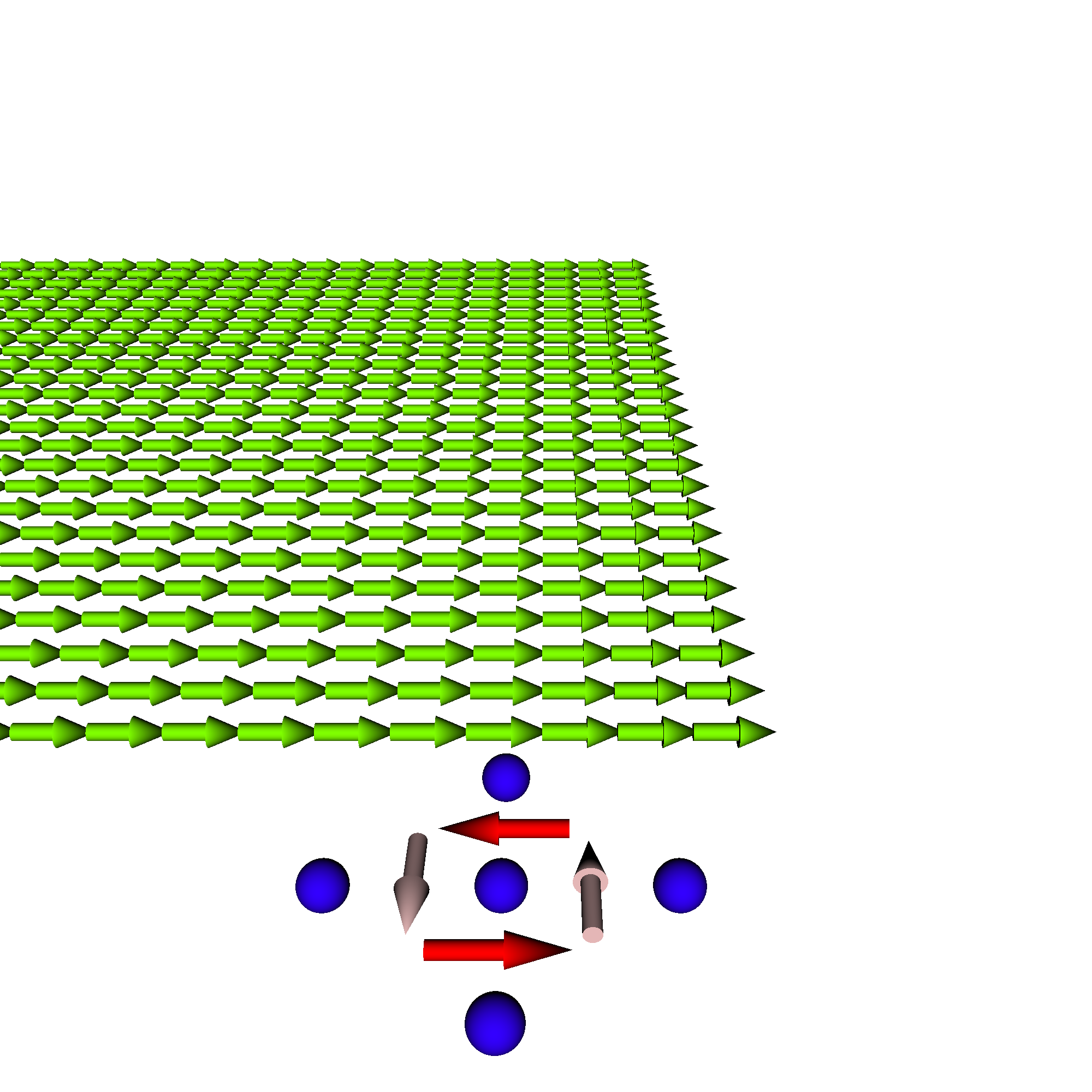} }
    \put( 0.0,7.6){ \makebox(\boxsize,\boxsize){(a)} } 
    \put( 2.0,7.6){ \makebox(4,\boxsize){Magnetization} } 
    \put( 1.7,2.2){ \makebox(\boxsize,\boxsize){(b)} } 
    \put( 5.3,0.9){ \makebox(2,\boxsize){DMI vectors} } 
    
    \put(7.5, 7.6){ \makebox(\boxsize,\boxsize){(c)} } 
    \put(7.5, 2.7){
      \put(0, 0){ \includegraphics[width=7.5 cm,trim={0 0 0 0},clip=true]{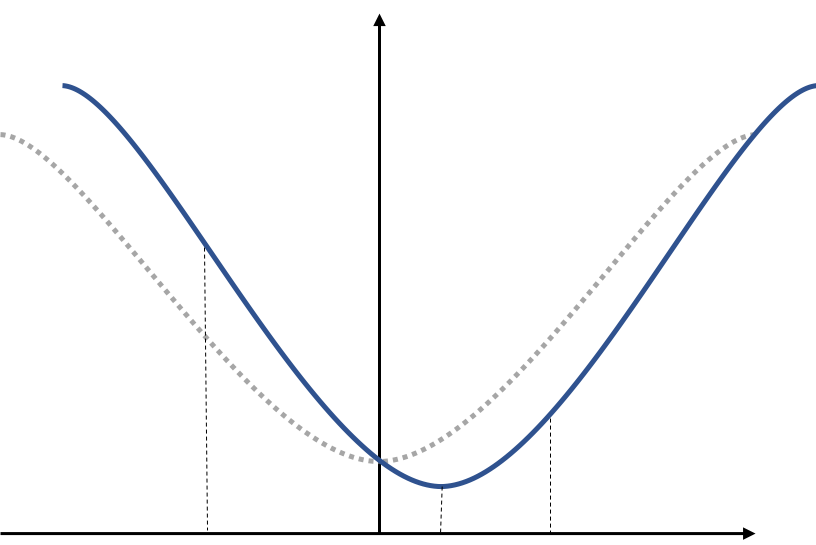} }

      \put(6.7, 0.3){ \makebox(\boxsize,\boxsize){$\VEC k$} } 
      \put(3.6, 4.7){ \makebox(\boxsize,\boxsize){$\omega$} } 

      \put(3.3,-0.3){ \makebox(\boxsize,\boxsize){$\Gamma$} } 
      \put(3.9,-0.3){ \makebox(\boxsize,\boxsize){$\delta \VEC k$} } 
      \put(4.9,-0.3){ \makebox(\boxsize,\boxsize){$\VEC q$} } 
      \put(1.6,-0.3){ \makebox(\boxsize,\boxsize){$-\VEC q$} } 
    }
  \end{picture}

  \caption{\label{fig:dispersion_shift_example}
  The shift of the spin-wave dispersion due to the DMI in ferromagnets.
  (a) In our convention, the magnetization direction of a ferromagnet is given by the direction of the spins.
  (b) Fragment of a square lattice showing the Dzyaloshinskii-Moriya-interaction vectors between the central atom and its neighbors, all lying in the plane.
  (c) The DMI components along the magnetization direction, shown in red in (b), induce an asymmetry of the spin-wave dispersion curve, which shifts sideways.
  Spin waves with opposite wavevectors, $\VEC q$ and $-\VEC q$, are no longer degenerate, such as in the absence of the DMI (indicated by the gray dotted line).
  Measuring the location of the new energy minimum $\delta \VEC k$ provides the chirality (spatial orientation) and magnitude of the DMI.
  }
\end{figure}

The theoretical realization that the spin-wave dispersion of ferromagnets can acquire an asymmetry due to the DMI was put forth by Udvardi and Szunyogh~\cite{udvardi_chiral_2009} and Costa \textit{et al.}~\cite{costa_spin-orbit_2010}.
The key requirement is that the magnetization and the DMI vectors are not perpendicular, which then leads to the nonreciprocity of the spin-wave dispersion (its energy minimum shifts away from the $\Gamma$--point), see Fig.~\ref{fig:dispersion_shift_example}. 
This means that the energies of spin waves with wavevectors of equal magnitude and opposite directions are no longer degenerate.
However, if the magnetization lies in a plane of mirror symmetry, the spin-wave dispersion remains reciprocal for wavevectors along the magnetization direction, as the effective DMI has to vanish in that case, due to Moriya's rules. 
Other authors have theoretically proposed to characterize the DMI from the spin-wave properties of thin films~\cite{moon_spin-wave_2013,cortes-ortuno_influence_2013}.
These seminal papers have opened a route to experimentally probe the DMI in ferromagnetic materials: the strength and chirality of the DMI can be deduced from the measured asymmetry of the spin-wave dispersion, for instance by fitting the data to a Heisenberg model Hamiltonian.
The chirality can be measured because it defines the direction in which the minimum of the spin-wave dispersion shifts away from the $\Gamma$--point.
It is worth mentioning that nonreciprocity without DMI has also been theoretically proposed for noncollinear magnetic structures~\cite{cheon_nonreciprocal_2018}.

There are different experimental techniques able to probe spin waves, such as inelastic scattering with electrons, neutrons or light, or broadband spectroscopy using coplanar waveguides, each with their capabilities and limitations.
In Ref.~\cite{zakeri_asymmetric_2010}, Zakeri \textit{et al.} used spin-polarized electron energy-loss-spectroscopy (SPEELS)~\cite{plihal_spin_1999,vollmer_spin-polarized_2003,michel_spin_2015,michel_high_2016,dos_santos_first-principles_2017} to experimentally detect the shift of the spin-wave dispersion due to the DMI in thin films of Fe/W(110).
The same principles have proven a fruitful way of accessing the DMI when applied to Brillouin light scattering experiments in thin film systems~\cite{cho_thickness_2015,di_direct_2015,nembach_linear_2015,belmeguenai_interfacial_2015,tacchi_interfacial_2017,camosi_anisotropic_2017} and to inelastic neutron scattering in bulk materials~\cite{sato_magnon_2016,weber_field_2018,weber_non-reciprocal_2018,weber_polarized_2019}, with broadband spectroscopy as alternative~\cite{schwarze_universal_2015,iguchi_nonreciprocal_2015,seki_magnetochiral_2016,kwon_giant_2016,lee_all-electrical_2016}.
Similarly, nonreciprocity was observed in antiferromagnets, but only when subjected to an external magnetic field~\cite{gitgeatpong_nonreciprocal_2017}.
We have recently proposed spin-resolved EELS (SREELS), which consists of a SPEELS setup augmented with a spin filter for the scattered electrons~\cite{dos_santos_spin-resolved_2018}.
Within SREELS, one has access to various spin-scattering channels, where the scattered electrons can either have their spins flipped or not.
In contrast to collinear magnets, where only spin-flip processes are responsible for the emission of spin waves, non-spin-flip processes can generate spin excitations in noncollinear materials~\cite{dos_santos_spin-resolved_2018}. 

In this work, we provide a complete characterization of the nonreciprocal effects in the spin-wave spectrum of complex magnetic structures.
We show that the angular momentum of a given spin-wave mode can be associated with its handedness -- a spatial chirality that defines the phase sign of precessing adjacent spin moments, which allows us to predict the effect of the DMI for that mode.
Furthermore, we demonstrate that only systems with finite total magnetization can feature a nonreciprocal total spin-wave spectrum, e.g., when considering the spin-wave energies of all modes.
Moreover, this nonreciprocity is observed on the reciprocal-space directions where the Fourier-transformed DMI vectors have finite projections on the magnetization.
In zero-net-magnetization systems, despite the lack of nonreciprocity of the total spin-wave spectrum, we uncover that individual spin-wave modes can be nonreciprocal.
These nonreciprocal modes usually come in pairs, each with opposite angular momentum leading to their dispersion curves to shift in opposite directions while keeping the total spin-wave spectrum reciprocal.
We also prove that spin-polarized experiments, such as SPEELS, SREELS, or polarized inelastic neutron scattering, can be used to reveal the DMI-induced nonreciprocity of individual spin-wave modes in noncollinear materials.
The nonreciprocity in practice leads to an asymmetric scattering rate for opposite wavevectors, which only appears when the probing-beam polarization aligns with a spin-wave angular momentum probing the Fourier-transformed-DMI components parallel to them.
Furthermore, we show that the angular momenta of the spin-wave modes are strongly related to the DMI, that is, they are given not only by the spin configuration, but they are also directly influenced by the DMI itself.
Thus, SREELS and SPEELS measurements would allow determining the chirality of the Dzyaloshinskii-Moriya interaction, which could be used to distinguish a  skyrmion from an antiskyrmions lattice, for example.


\section{Theoretical framework and model systems}
\label{sec:theoretical_framework}
    
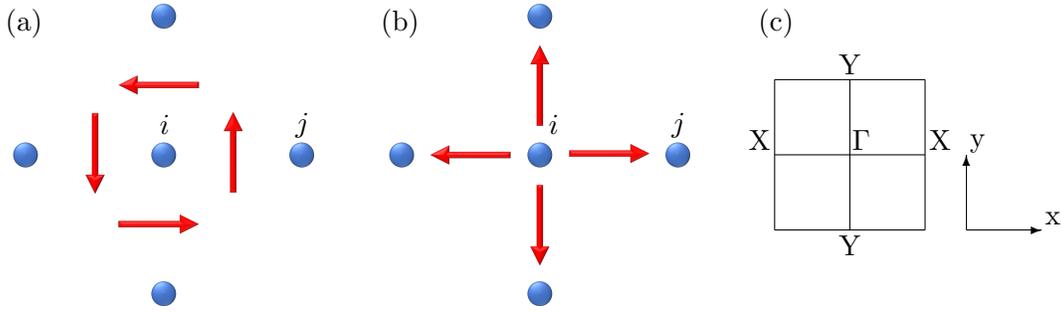
\begin{figure}[t]
\setlength{\unitlength}{1cm} 
\newcommand{\boxsize}{0.3}

  \begin{picture}(15,4.5)
    \put(0.5, 0){ \includegraphics[width=4.0 cm,trim={0 0 0 0},clip=true]{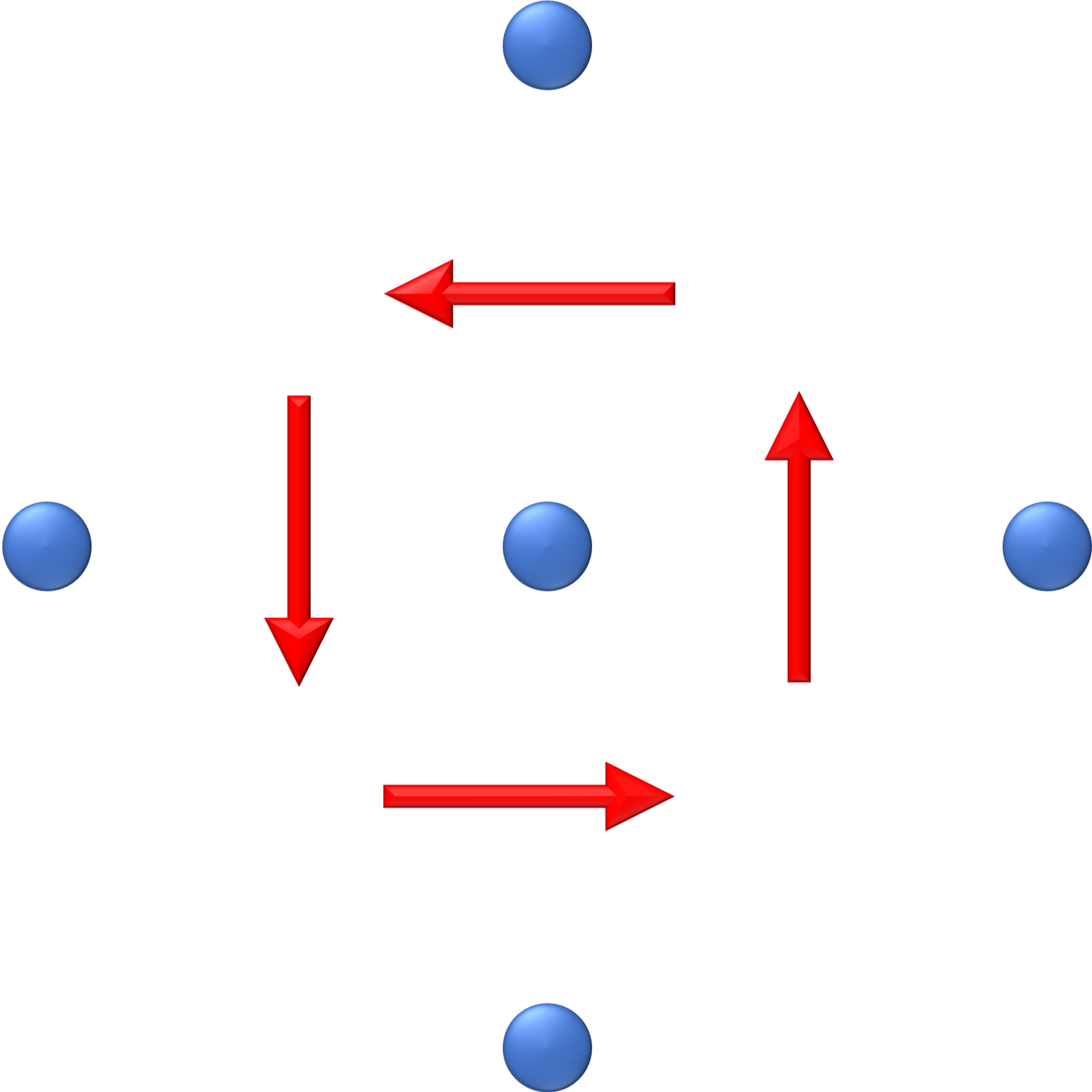} }
    \put( 2.37,2.25){ \makebox(\boxsize,\boxsize){$i$} } 
    \put( 4.20,2.25){ \makebox(\boxsize,\boxsize){$j$} } 
    
    \put(5.5, 0){ \includegraphics[width=4.0 cm,trim={0 0 0 0},clip=true]{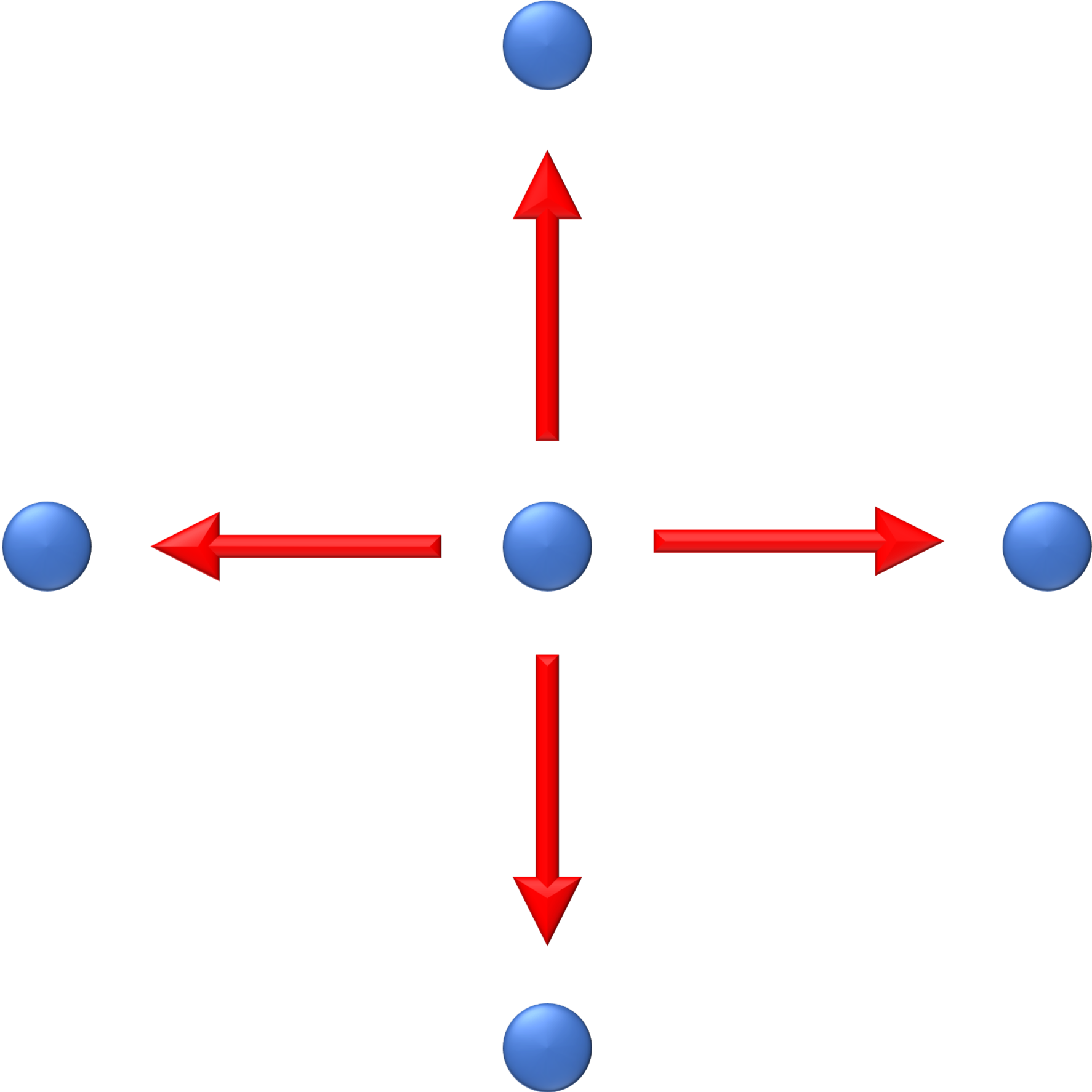} }
    \put( 7.54,2.25){ \makebox(\boxsize,\boxsize){$i$} } 
    \put( 9.20,2.25){ \makebox(\boxsize,\boxsize){$j$} } 

    \put( 0.5,3.6){ \makebox(\boxsize,\boxsize){(a)} } 
    \put( 5.5,3.6){ \makebox(\boxsize,\boxsize){(b)} } 
    \put(10.5,3.6){ \makebox(\boxsize,\boxsize){(c)} } 

    \put(10.5, 1.0){
      \scalebox{1.0}{
        \put(0,0){ \line(1,0){2}}
        \put(0,1){ \line(1,0){2}}
        \put(0,2){ \line(1,0){2}}

        \put(0,0){ \line(0,1){2}}
        \put(1,0){ \line(0,1){2}}
        \put(2,0){ \line(0,1){2}}

        \put(-0.35, 1.05){ \makebox(\boxsize,\boxsize){X} } 
        \put( 2.05, 1.05){ \makebox(\boxsize,\boxsize){X} } 
        \put( 0.85,-0.35){ \makebox(\boxsize,\boxsize){Y} }     
        \put( 0.85, 2.05){ \makebox(\boxsize,\boxsize){Y} }     
        \put(    1, 1.05){ \makebox(\boxsize,\boxsize){$\Gamma$} }     

        \put(2.55, 0){
          \put(0,0){ \vector(1,0){1}}
          \put(0,0){ \vector(0,1){1}}
          \put(1.0,0.0){ \makebox(\boxsize,\boxsize){x} } 
          \put(0.0,1.0){ \makebox(\boxsize,\boxsize){y} }     
        }
      }
    }
  \end{picture}

  \caption{\label{fig:models}
  The two model systems considered in this work.
  Both consist of a square lattice with nearest-neighbor interactions only.
  The exchange interaction is the same for both models, but (a) Model I has DMI vectors perpendicular to the bonds and swirling counterclockwise, while (b) Model II has DMI vectors diverging from the sites being parallel to the bonds.
  Model I has a cycloidal spin spiral as its ground state, while Model II realizes a helical spiral.
  (c) Brillouin zone with its high-symmetry points and our choice of the frame of reference: The Y--$\Gamma$--Y path is along $\hat{\VEC y}$ and X--$\Gamma$--X is along $\hat{\VEC x}$. 
  }
\end{figure}

To clarify the interplay between the DMI, the ground-state magnetic structure and the properties of its spin-wave spectrum, we adopt two very simple spin models that allow us to explore all of the involved aspects.
These are based on the following generalized classical Heisenberg model, whose Hamiltonian reads 
\begin{equation} \label{eq:HeisenbergHamiltonian}
  \HH = -\frac{1}{2} \sum_{ij} \left( J_{ij} \VEC S_i \cdot \VEC S_j + \VEC D_{ij} \cdot \VEC S_i \times \VEC S_j \right) - \sum_{i} \VEC B \cdot \VEC S_i \quad ,
\end{equation}
where $J_{ij}$ is the magnetic exchange interaction parameter, $\VEC D_{ij}$ is the Dzya\-lo\-shi\-skii-Moriya interaction vector between sites $i$ and $j$, and $\VEC B$ is a uniform external magnetic field.
We take a square lattice (lattice constant $a$) for both models, with the magnetic interactions restricted to nearest-neighbors.
The $J_{ij}$ are identical in both models ($J$ for all nearest-neighbors), but the set of $\VEC D_{ij}$ vectors differs (note that $\VEC D_{ji} = -\VEC D_{ij}$):
Model I has the DMI vectors perpendicular to the bond connecting the corresponding sites, lying in the plane of the lattice and swirling counterclockwise, see Fig.~\ref{fig:models}(a);
Model II has the DMI vectors parallel to the bonds, also lying in-plane and radiating outwards from the site $i$ to its neighbors, see Fig.~\ref{fig:models}(b).
Figure~\ref{fig:models}(c) shows the square-lattice Brillouin zone marking its high-symmetry points.
For the simulation presented throughout the paper, we took the parameters to be: $J=1$, $D=1$.

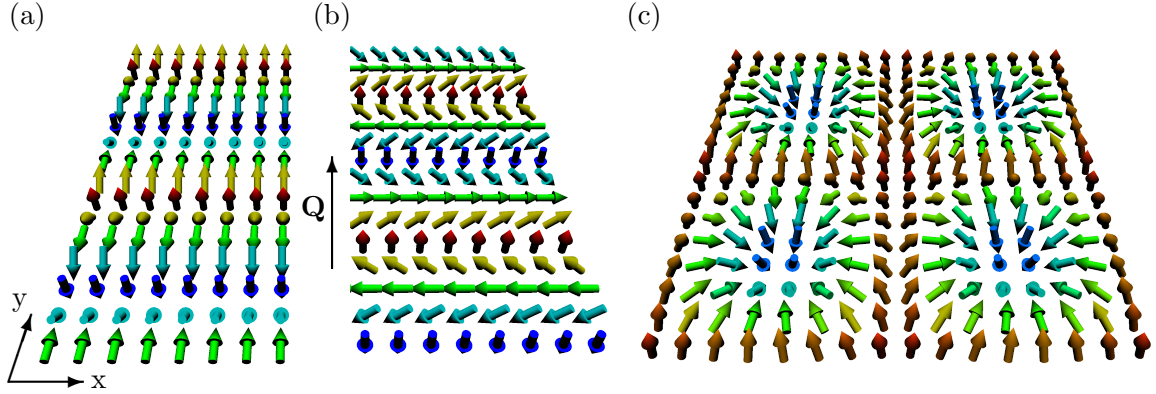
\begin{figure}[t]
\setlength{\unitlength}{1cm} 
\newcommand{\boxsize}{0.3}

  \begin{picture}(15,5)
    \put( 0, 0){ \includegraphics[width=3.5 cm,trim={0 7cm 35.6cm 15cm},clip=true]{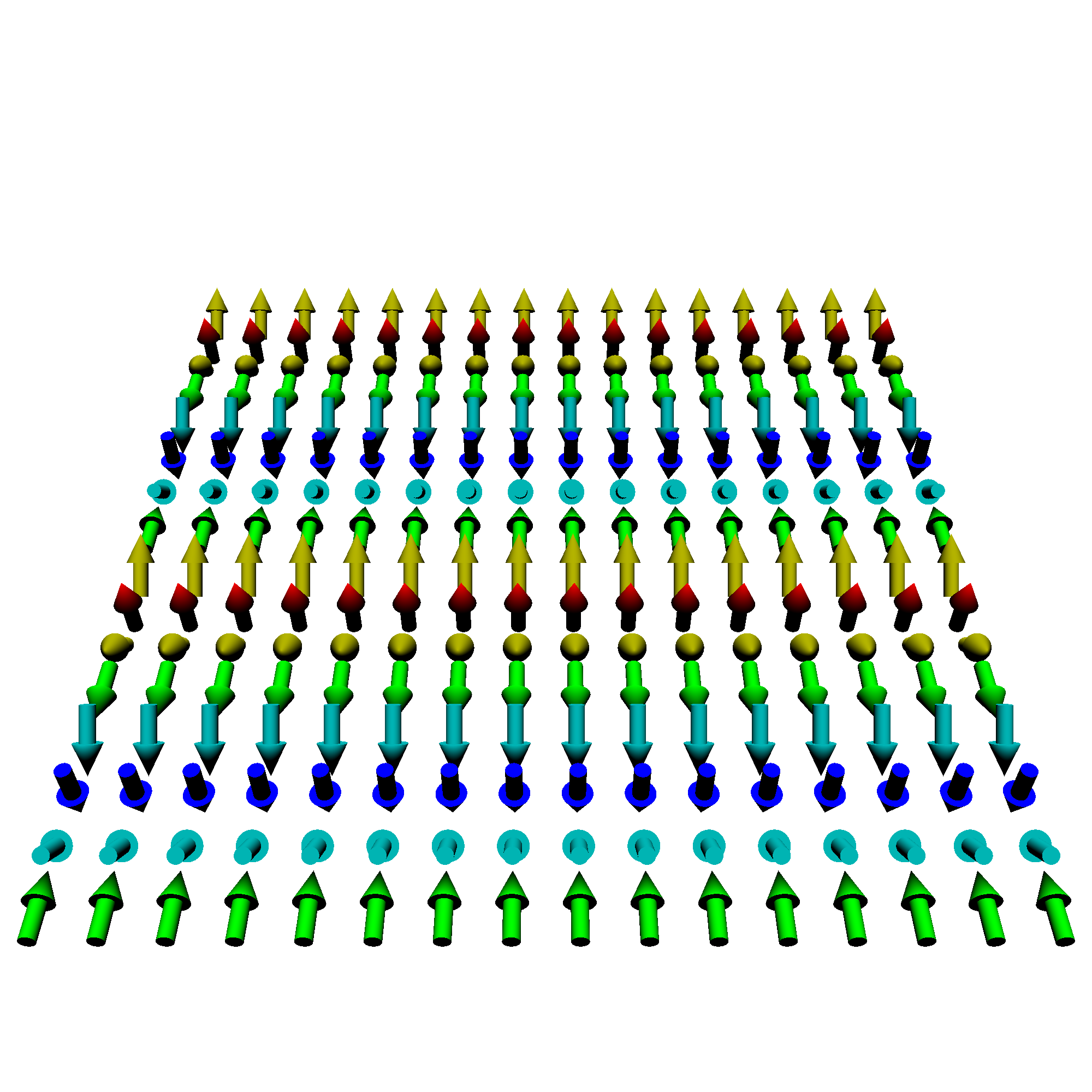} }
    \put( 4.25, 0){ \includegraphics[width=3.5 cm,trim={35.6cm 7cm 0 15cm},clip=true]{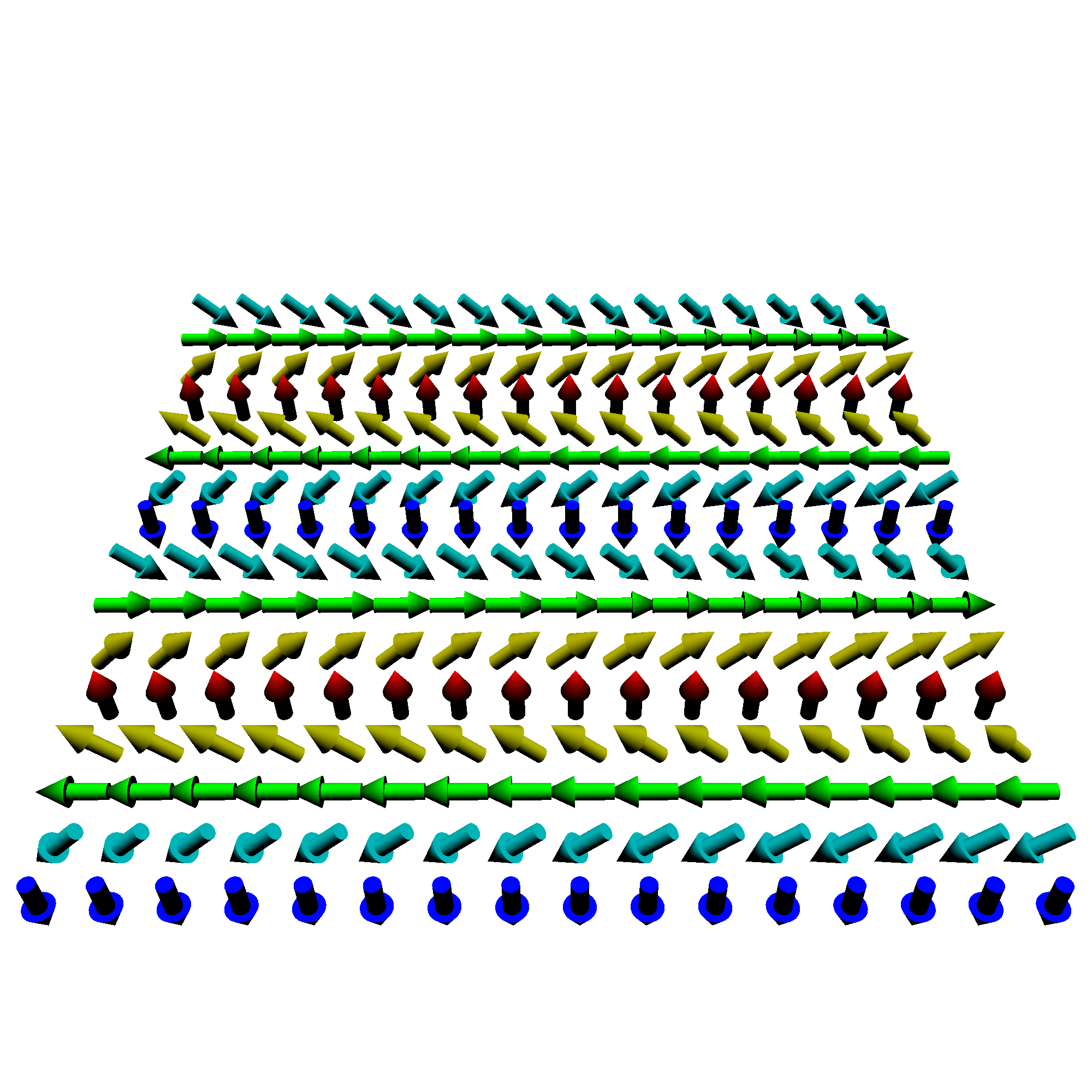} }
    
    \put( -0.2 , 4.7){ \makebox(\boxsize,\boxsize){(a)} } 
    \put(  3.85, 4.7){ \makebox(\boxsize,\boxsize){(b)} } 

    \put(-0.3, 0){
        \thicklines
        \put(0,0){ \vector(1,0){1}}
        \put(0,0){ \vector(1,3){0.3}}
        \put(1.05,-0.15){ \makebox(\boxsize,\boxsize){x} } 
        \put( 0.0, 0.9 ){ \makebox(\boxsize,\boxsize){y} }     
    }
    
    \put(3.60,1.5){
        \thicklines
        \put(0.4,0){ \vector(0,1){1.5}}
        \put(0,0.65){ \makebox(\boxsize,\boxsize){$\VEC Q$} }
    }
    
    \put(8, 0){ \includegraphics[width=7 cm,trim={0 7cm 0 15cm},clip=true]{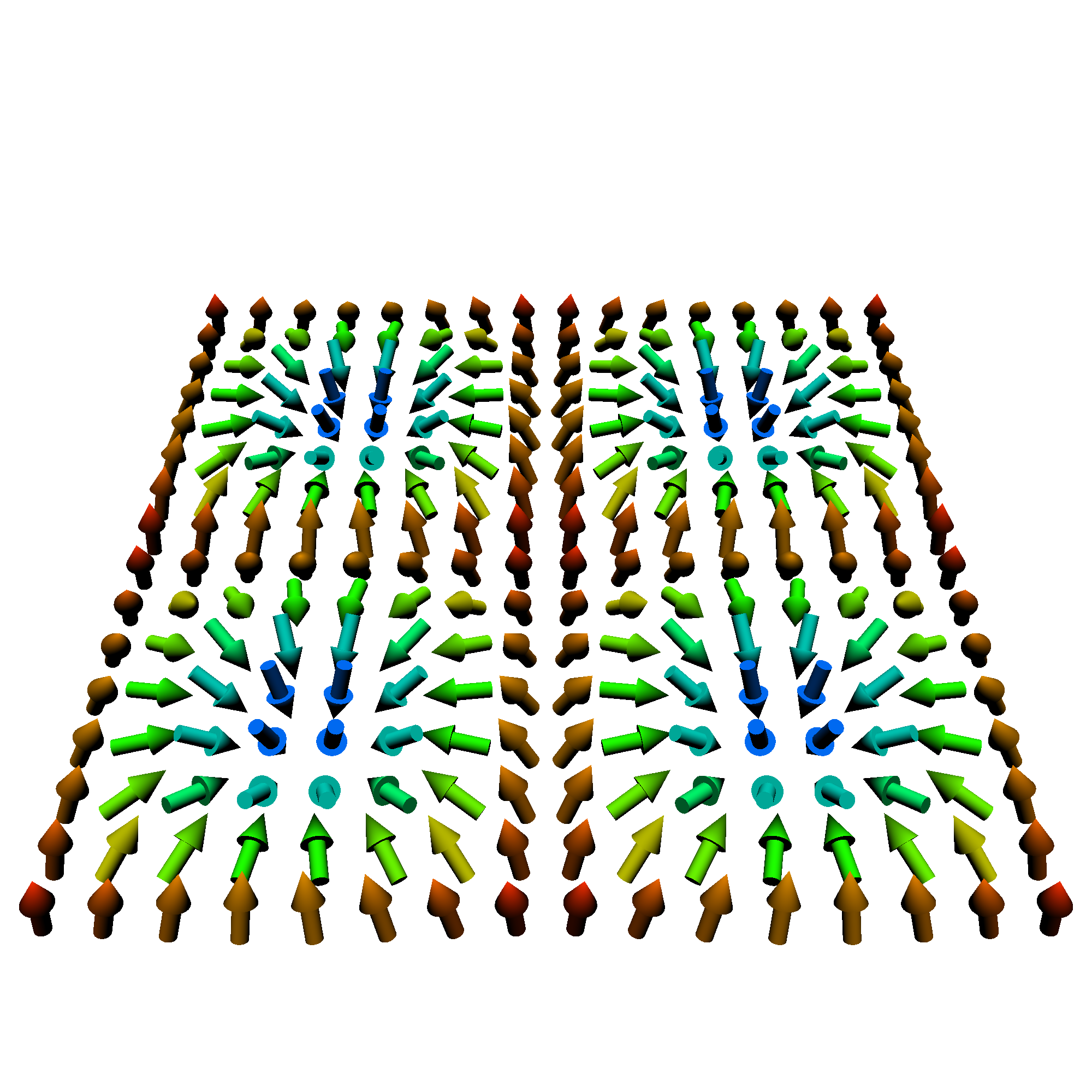} }
    \put(8, 4.7){ \makebox(\boxsize,\boxsize){(c)} } 
  \end{picture}

  \caption{\label{fig:spin_configurations}
  Spin configuration stabilized by the two models, which assume the MEI and DMI to be limited to the nearest neighbors and $J=D=1$.
  (a) A cycloidal spin spiral being the ground state of Model I.
  (b) The helical spin spiral stabilized by Model II.
  Both spin spiral have the same wavevector $\VEC Q = \frac{2\pi}{8a}\,\hat{\VEC y}$.
  (c) Skyrmion lattice obtained by adding an out-of-plane magnetic field to Model I.
  }
\end{figure}

We find the ground-state spin configuration for Models I and II using atomistic spin dynamics simulations by solving the Landau–Lifshitz–Gilbert equation with the \textit{Spirit} code~\cite{muller_spirit:_2019}.
Using a unit cell of $8 \times 8$ atoms, one obtains for Model I a cycloidal spin spiral, see Fig.~\ref{fig:spin_configurations}~(a), and a helical spin spiral for Model II, Fig.~\ref{fig:spin_configurations}~(b).
In these figures, the wavevector $\VEC Q = (2\pi / 8) \hat{\VEC y}$ of the spin spirals is along $\hat{\VEC y}$, however, the spin spirals with wavevector along $\hat{\VEC x}$ are also possible, which are degenerate to the ones we are showing.
By adding an external magnetic field normal to the film in Model I, we can stabilize a skyrmion lattice as shown in Fig.~\ref{fig:spin_configurations}~(c).
In this case, the square arrangement of skyrmions is imposed by the choice of the unit cell.
The direction of the net magnetization of any spin texture will be denoted by $\VEC n^0$.

The spin-wave modes and the corresponding spin-wave spectrum is computed with our theory for SREELS of noncollinear magnets discussed in Ref.~\cite{dos_santos_spin-resolved_2018}.
In a nutshell, we employ time-dependent perturbation theory to describe the interaction between the probing beam and the magnetic system.
The spin-wave excitations are computed out of the self-consistently-determined classical ground state, in the adiabatic approximation.
The unfolding of the spin-wave modes and the potential extinction of their signal due to destructive interference are naturally taken into account by this theory.

SREELS provides spin-resolved spectroscopy of the spin waves.
In this setup, a spin-polarized beam of electrons is used to probe the magnetic material (this could be changed to neutrons with little modification).
The scattered electrons are then spin-filtered with the spin analyzer collinear with the incident beam polarization.
This gives rise to four scattering channels, one for each possible combinations of [incoming spin]-[outgoing spin].
Two of these channels correspond to non-spin-flip processes, namely the up-up and the down-down channels.
The other two, up-down and down-up, account for spin-flip events, where angular momentum is exchanged with the sample.

When probing a ferromagnet with all spins along $+z$, only the down-up channel can excite spin waves (assuming the probing beam polarization to be parallel to the ferromagnetic magnetization), because this process transfers the exact angular momentum required to excite a quantum of spin wave (the net angular momentum of the spin wave is $-1$ in units of $\hbar$).
In contrast, a spin spiral hosts three types of spin-wave modes (also known as `universal helimagnon modes'~\cite{schwarze_universal_2015}).
If the beam polarization is aligned perpendicular to the plane where the magnetic moments rotate in the ground state, their net angular momentum can be inferred from the spin-angular-momentum conservation that defines the four scattering channels in SREELS~\cite{dos_santos_spin-resolved_2018}.
One mode appears in the up-down channel and another in the down-up channel, so these are rotational modes with the net angular momentum of $+1$ and $-1$, respectively.
The third type of mode appears in the up-up and down-down channels, and so has zero net angular momentum.
If the beam polarization is not set as explained, different types of modes can be detected in the same scattering channel.


\section{Results and discussions} 
\label{sec:results}

The following summarizes how the Dzyaloshinskii-Moriya interaction affects the dynamics and energetics of spin waves in collinear and noncollinear magnetic structures followed by extended discussions in the next subsections: 
\begin{description}
  \item[(A)] Nonreciprocal spin-wave spectrum only occurs, in the absence of an external magnetic field, for systems of finite magnetization and when $\VEC n^0 \cdot \VEC D(\VEC k) \neq 0$, i.e., if the projection of the Fourier-transformed DMI on the magnetization direction is finite.
  \item[(B)] The angular momentum of a spin-wave mode can be regarded as the handedness attribute, which defines the direction towards which the dispersion of the given mode shifts out of the $\Gamma$--point due to the DMI. 
  \item[(C)] Systems of zero net magnetization can host spin-wave modes individually nonreciprocal induced by the DMI, while the total spin-wave spectrum remains reciprocal. An external magnetic field can induce nonreciprocity.
  \item[(D)] Polarized inelastic-scattering experiments can be used to unveil the DMI-induced nonreciprocity, and thus allowing to measure the DMI orientation. A nonreciprocal spectrum only occurs for spin-flip scattering processes due to spin-wave modes whose angular momentum aligns with the polarization of the probing particles and $\VEC D(\VEC k)$.  
  \item[(E)] All spin textures that are favored by the DMI have nonreciprocal spin-wave modes with angular momentum aligned to the component of $\VEC D(\VEC k)$ that contributes to the DMI energy gain.
\end{description}

\subsection{Nonreciprocal spin-wave spectrum} 
\label{sub:nonreciprocal_spin_wave_spectrum}

In the absence of an external magnetic field, a nonreciprocal spin-wave spectrum (different spin-wave energies for modes with wavevectors which are equal in length and opposite in direction) only occurs for systems with finite magnetization.
Such a nonreciprocity manifests in the reciprocal-space directions along which a component of $\VEC D(\VEC k)$ aligns with the net magnetization.

The first statement is related to the breaking of time-reversal symmetry.
Consider a system described by the Hamiltonian $H$.
If a system is invariant under time reversal operator $\Theta$, then $\Theta H(\VEC k) \Theta^{-1} = H(-\VEC k)$,
and the reciprocity of the system is guaranteed.
Systems of zero net magnetization, such as antiferromagnets and some spin spirals ( e.g., see Figs.~\ref{fig:spin_configurations}(a-b)),
are not invariant under time reversal, nor under partial translation $T_{\lambda/2}$ (translation by half of the spin spiral wavelength $\lambda$ along the spiral propagation direction), individually.
However, they are invariant under a combined operation of time reversal plus partial translation $\mathcal S = \Theta T_{\lambda/2}$, which leads to $\mathcal S_{\VEC k} H(\VEC k) \mathcal S_{\VEC k}^{-1} = H(-\VEC k)$ \cite{mong_antiferromagnetic_2010}.
When the system has a finite net magnetization, it is not possible to find such a combined operation that leaves the Hamiltonian invariant.

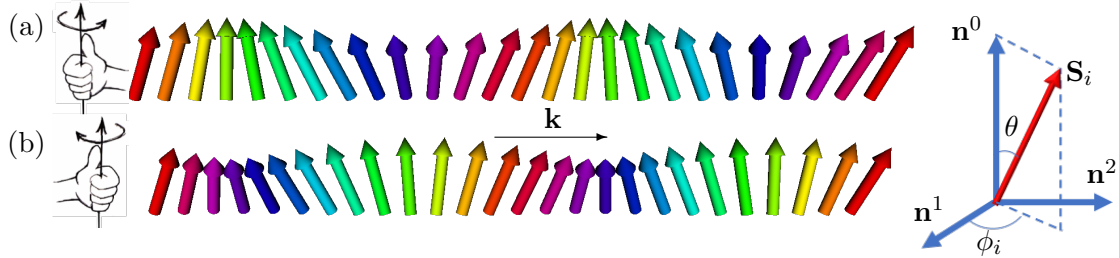
\begin{figure}[t]
\setlength{\unitlength}{1cm} 
\newcommand{\boxsize}{0.3}

  \begin{picture}(15,3.5)

    \put( 0.5, 0.5){ 
      \put( 0.4, 0.0){ \includegraphics[width=11.3 cm,trim={0cm 22cm 0cm 17cm},clip=true]{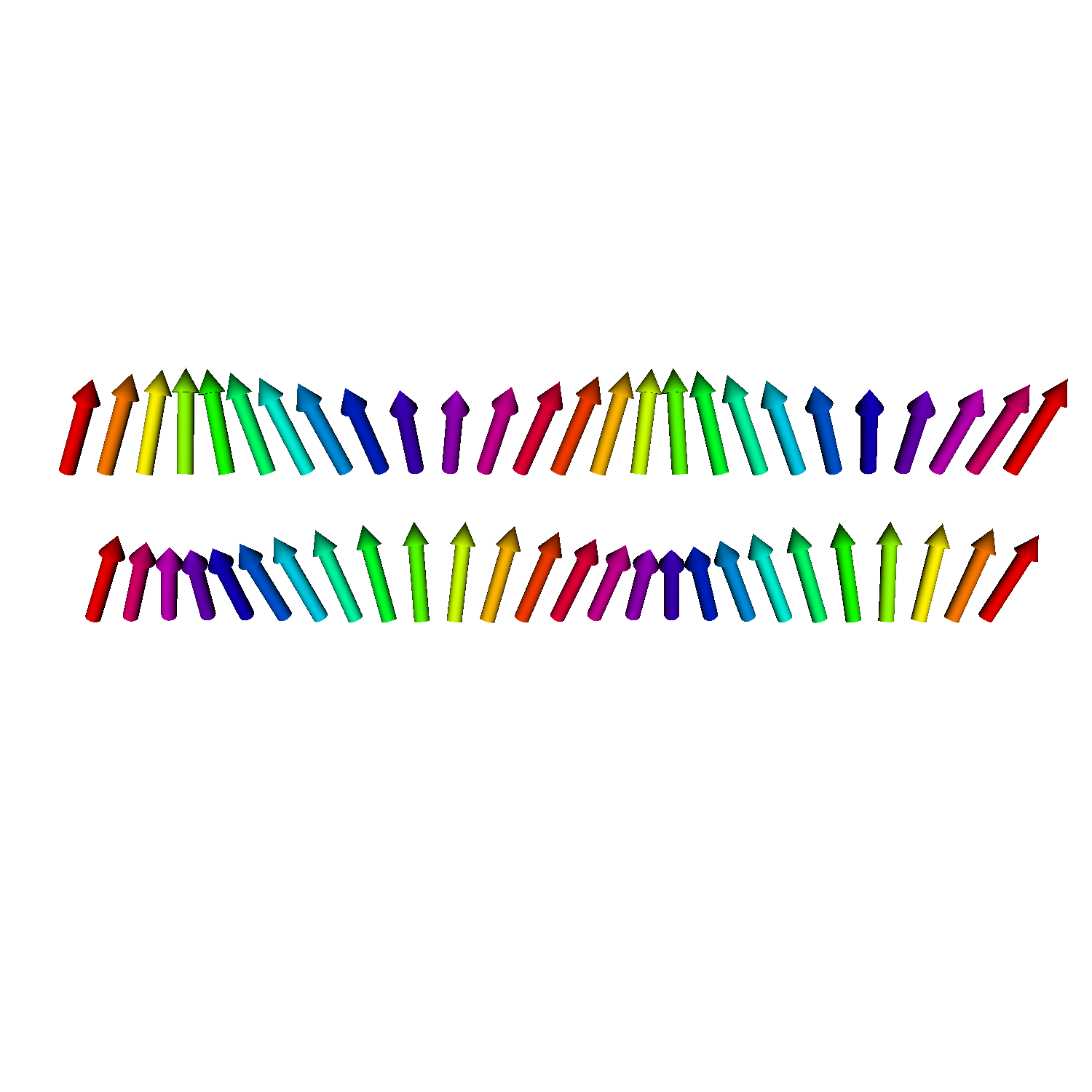} }
      \put( 0, 1.5){ \includegraphics[width=1 cm,trim={0 0 0 0},clip=true]{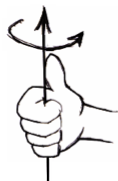} }
      \put( 0, 0){ \includegraphics[width=1 cm,trim={0 0 0 0},clip=true]{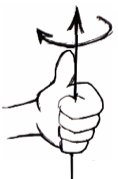} }

      \put(1, 0){
        \put( 5.5, 1.3){ \makebox(\boxsize,\boxsize){$\VEC k$} }
        \put( 5.0, 1.2){\vector(1,0){1.5}}
      }

      \put( -0.5, 2.5){ \makebox(\boxsize,\boxsize){(a)} }
      \put( -0.5, 0.95){ \makebox(\boxsize,\boxsize){(b)} }
    }

    \put(11.8, 0) {
      \put( 0, 0.0){ \includegraphics[width=3 cm,trim={0 0 0 0},clip=true]{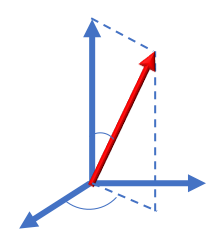} }
      \put( 0.7, 3.0){ \makebox(\boxsize,\boxsize){$\VEC n^0$} } 
      \put( 0.2, 0.6){ \makebox(\boxsize,\boxsize){$\VEC n^1$} } 
      \put( 2.5, 1.1){ \makebox(\boxsize,\boxsize){$\VEC n^2$} } 
      \put( 1.3, 1.7){ \makebox(\boxsize,\boxsize){$\theta$} } 
      \put( 1.0, 0.15){ \makebox(\boxsize,\boxsize){$\phi_i$} } 
      \put( 2.2, 2.4){ \makebox(\boxsize,\boxsize){$\VEC S_i$} } 
    }
  \end{picture}

  \caption{\label{fig:snapshot_spinwave}
  Instantaneous snapshot of the spin wave of a ferromagnet for a given wavevector $\VEC k$ as given by Eq.~\eqref{eq:spin_wave_snapshot}.
  The chirality is defined as the sense in which the spin moments rotation as we proceed along the propagation direction given by $\VEC{k}$.
  (a) For $c=+1$, the spin wave has a right-handed chirality.
  (b) For $c=-1$, it has a left-handed chirality.
  The magnetization direction is given by $\VEC n^0$.
  During the precession due to the spin wave, all spins deviate from $\VEC n^0$ by a fixed angle $\theta$.
  The phase of precession of the $i$--th spin is given by $\phi_i = \VEC k \cdot \VEC R_i$, where $\VEC R_i$ is the spin position, and it is used to color code the spins.
  }
\end{figure}

We can prove the second statement, following Ref.~\cite{udvardi_chiral_2009}, considering an instantaneous snapshot of a classical spin wave in a ferromagnet, given by  
\begin{equation} \label{eq:spin_wave_snapshot}
\begin{split}
  \VEC S_i = \cos\phi_i \sin\theta\,\VEC{n}^1 + c \sin\phi_i \sin\theta\,\VEC{n}^2 + \cos\theta\,\VEC{n}^0  \quad , \\
\end{split}
\end{equation}
where $\VEC n_0$ is a unit vector along the magnetization, which forms an orthonormal basis together with $\VEC n^1$ and $\VEC n^2$, see Fig.~\ref{fig:snapshot_spinwave}.
$\theta$ corresponds to a small deviation from the magnetization direction $\VEC n^0$, while $\phi_i = \VEC k \cdot \VEC R_i$ corresponds to a transversal rotation of the spin moments with rotational sense (chirality) given by $c = \pm 1$.
Placing this expression into Eq.~\eqref{eq:HeisenbergHamiltonian}, we obtain that the only chirality-dependent term is given by
\begin{equation}\label{eq:FM_chiral_asymmetry}
  E(\VEC k, c) \propto c ~ \VEC n^0 \cdot \VEC D(\VEC k)  \quad ,
\end{equation}
where $\VEC D(\VEC k)$ is the lattice Fourier transform of the Dzyaloshinskii-Moriya vectors.
For details on how to obtain the above equation, see Appendix~\ref{Apxsec:appendix_spin_wave_chirality_in_ferromagnets}.

\begin{figure}[t]
\setlength{\unitlength}{1cm} 
\newcommand{\boxsize}{0.3}

  \begin{picture}(15,7.5)
    \put(0, 0){ \includegraphics[width=14.5 cm,trim={0 0 0 0},clip=true]{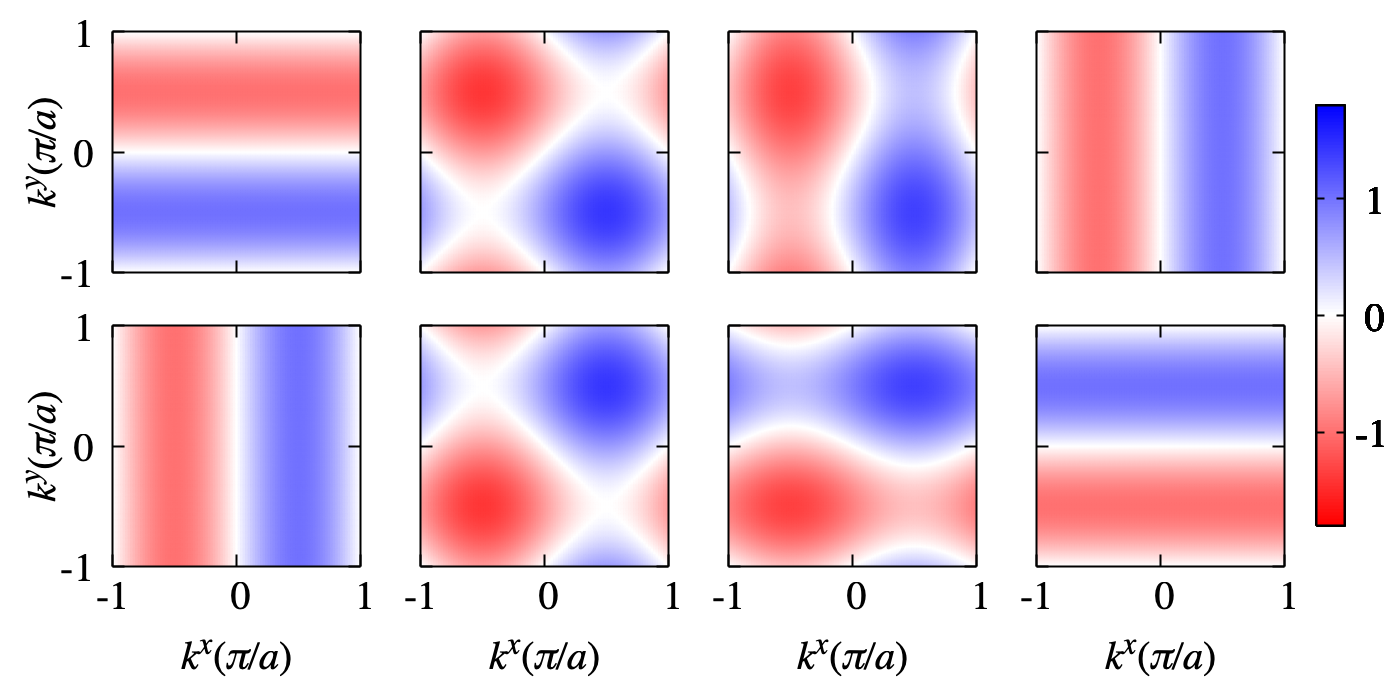} }

    \put(0.0,6.7){ \makebox(\boxsize,\boxsize){(a)} } 
    \put(0.0,3.6){ \makebox(\boxsize,\boxsize){(b)} } 
    \put(13.75,6.35){ \makebox(\boxsize,\boxsize){$E(\VEC k)$} } 

    \thicklines
    \put(0, 5.68){
      \put( 2.45, 0){ \vector(1,0){1}}
      \put( 5.65, 0){ \vector(1,1){0.70710678118655}}
      \put( 8.83, 0){ \vector(1,2){0.44721359549996}}
      \put(12.02, 0){ \vector(0,1){1}}
    }
    \put(0, 2.63){
      \put( 2.45, 0){ \vector(1,0){1}}
      \put( 5.65, 0){ \vector(1,1){0.70710678118655}}
      \put( 8.83, 0){ \vector(1,2){0.44721359549996}}
      \put(12.02, 0){ \vector(0,1){1}}
    }
  \end{picture}

  \caption{\label{fig:FM_chiral_asymmetry}
  Chirality-dependent spin-wave energy landscape throughout the Brillouin zone, obtained from Eq.~\eqref{eq:FM_chiral_asymmetry}.
  The (a) row corresponds to the energy landscape for Model I and row (b) for Model II.
  Each column corresponds to a different in-plane magnetization direction, which is represented by the black arrows.
  }
\end{figure}
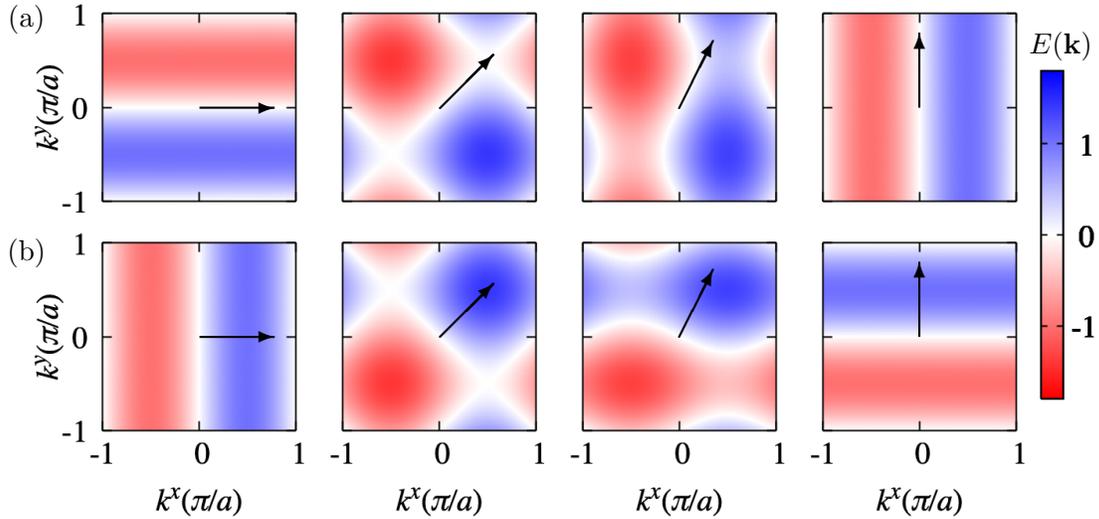

In both models, the ferromagnetic state can be stabilized by an external magnetic field, but the chiral asymmetry will only manifest when the magnetization has a finite in-plane projection.
For Model I, the Fourier transformation of the DMI interaction gives
$\VEC D(\VEC k) = 2 D\left(-\sin( a k^y )\,\hat{\VEC x} + \sin( a k^x )\,\hat{\VEC y}\right)$, and therefore, the asymmetry is strongest for spin waves propagating perpendicularly to the magnetization, and mostly vanishes when parallel to it, see Fig.~\ref{fig:FM_chiral_asymmetry}~(a).
For Model II, however, $\VEC D(\VEC k) = 2 D \left(\sin( a k^x )\,\hat{\VEC x} + \sin( a k^y )\,\hat{\VEC y}\right)$ and the asymmetry is strongest mostly for wavevectors parallel to the magnetization, see Fig.~\ref{fig:FM_chiral_asymmetry}~(b).


\subsection{Spin-wave angular momentum and spin-wave handedness}
\label{sub:spin_wave_angular_momentum_and_its_handedness}

Now we need to establish an important relation between spin-wave chirality, handedness and angular momentum.
In the previous section, our \textit{ansatz} of spin waves considers two possible spin-wave chiralities.
In the following, we demonstrate that only one of them is a solution to the coupled equation of motions that govern the dynamics.
Besides that, we define a spin-wave handedness, which is a chiral invariant for the spin waves whose sign is related to the direction of the spin-wave dispersion shift in the reciprocal space.
Lastly, we show that there is a one-to-one relation between the spin-wave handedness and the angular momentum.
That relation is fundamental in providing an easy and comprehensive way to predict chiral asymmetry in spin-wave dynamics induced by DMI. 

Thus far, we know that the spin-wave dispersion curve of a ferromagnet can be shifted out of the $\Gamma$--point due to the influence of the Dzyaloshinskii-Moriya interactions.
This shift was measured in the electron scattering experiments of Zakeri \textit{et al.} \cite{zakeri_asymmetric_2010}, and it occurs towards a very well-defined direction for a fixed direction of the magnetization (given that the DMI is a constant of the material).
From this fact, we can infer that spin waves in a ferromagnet have a given handedness that defines how the spin-wave energies respond to the DMI, for example, setting the direction of the dispersion shift.
Can a ferromagnet of fixed magnetization host spin waves of opposite handedness, such that their dispersion curves would shift to the opposite directions?
A hint comes from the fact that spin waves in a ferromagnetic system always possess angular momenta along the same direction (antiparallel to the magnetization~\footnote{The magnetization, which is the volumetric density of magnetic moment, is antiparallel to spin angular momentum because of the negative electric charge of the electrons. In the literature, however, often the minus sign is disregarded, which is the convention we follow in this paper.}).

With the previous question in mind, we will review the motion of the spin moments of a ferromagnet when hosting a spin wave.
We consider classical spin moments represented by vectors and the phenomenological Landau-Lifshitz equation describing the time evolution of every spin moment:
\begin{equation}\label{eq:eq_motion}
  \dv{\VEC S_i(t)}{t} = - \gamma \VEC S_i(t) \times \VEC B_i^{\text{eff}}(t) \quad ,
\end{equation}
where $\gamma$ is the  gyromagnetic ratio.
The effective field is given by
\begin{equation}\label{eq:eff_field}
  \VEC B_i^{\text{eff}}(t) = - \partdv{H}{\VEC S_i} =\sum_{j} \left(  J_{ij} \VEC S_j +   \VEC S_j \times \VEC D_{ij}   \right) + \VEC B_i \quad,
\end{equation}
where we considered the Hamiltonian of Eq.~\eqref{eq:HeisenbergHamiltonian}.
We have one equation of motion for each magnetic atom of our material, and these equations are coupled because the effective field in each site depends on the dynamics of the neighboring site to which they couple to via the magnetic interactions.

The presence of the DMI can cause instability in the ferromagnetic phase in favor of the spin-spiral structure.
To avoid this problem, we apply a sufficiently large external magnetic field along the $z$ direction.
We assume that the spin precession is of small amplitude around its equilibrium direction.
The solution steps for the linearized problem are collected in Appendix.~\ref{Apxsec:appendix_spin_wave_chirality_in_ferromagnets}.
The time evolution of the spin at site $i$ reads
\begin{equation}\label{eq:spin_wave}
\begin{split}
  \VEC S_i(\VEC k,t) =&
   \frac{ 1 }{\sqrt N} \big( 
   \cos(-\VEC k \cdot \VEC R_i + \omega_\VEC k t)\,\hat{\VEC x}+
   \sin(-\VEC k \cdot \VEC R_i + \omega_\VEC k t)\,\hat{\VEC y} \big)
   + S\,\hat{\VEC z} \quad ,
\end{split}
\end{equation}
which corresponds to a spin wave of wavevector $\VEC k$.
Its frequency $\omega_\VEC k$ is given by
\begin{equation}\label{eq:frequency}
  \omega_{\VEC k} =  S  \left( J_{\VEC 0} - J_{\VEC k}^+ \right) + B \quad \textnormal{with} \quad 
 J_{\VEC k}  = \sum_i A_{ij} \cos \big( \VEC k \cdot \VEC R_{ij} + \phi_{ij} \big) \quad ,
\end{equation}
where $A_{ij} = \sqrt{(D^z_{ij})^2 + J_{ij}^2 }$ and $\phi_{ij} = \arctan\big( D_{ij}^z/J_{ij} \big)$.
We have that $\omega_\VEC k \geq 0$ and thus every spin has a counterclockwise precession around the magnetization.
In the ferromagnetic ground state, all the spins are aligned, and so the total angular momentum of the system is maximal along $\hat{\VEC z}$ (the magnetization direction). 
With a spin wave, as the spins are precessing, the total angular momentum is reduced, which means that the spin-wave angular momentum is antiparallel to $\hat{\VEC z}$.

\begin{figure}[t]
\setlength{\unitlength}{1cm} 
\newcommand{\boxsize}{0.3}

  \begin{picture}(15,5)
    \put(0,1.2){
      \put(0, 2){ \includegraphics[width=14.5 cm,trim={0 0 0 0},clip=true]{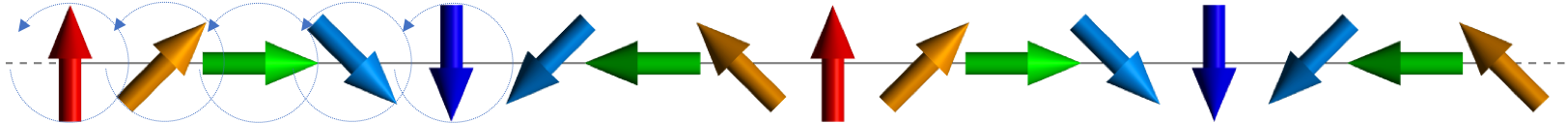} }

      \put(0.0,3.2){ \makebox(\boxsize,\boxsize){(a)} } 

      \put( 0.3,1.6){
        \put(7.3,0){ \makebox(\boxsize,\boxsize){$\VEC S_1$} }
        \put(8.2,0){ \makebox(\boxsize,\boxsize){$\VEC S_2$} }
      }

      \put(7.9,3.3){ \makebox(\boxsize,\boxsize){ $\hat{\VEC r}_{12}$ } }
      \put(7.6,3.2){ \vector(1,0){0.9}}

      \put(9.9,3.3){ \makebox(\boxsize,\boxsize){$\VEC k$} }
      \put(9.5,3.2){ \vector(1,0){1.2}}
    }

    \put(0,0.4){
      \put(0, 0){ \includegraphics[width=14.5 cm,trim={0 0 0 0},clip=true]{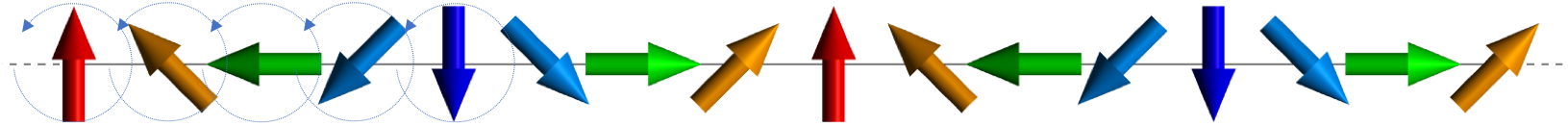} }

      \put(0.0,1.2){ \makebox(\boxsize,\boxsize){(b)} } 

      \put( 0.3,-0.4){
        \put(7.3,0.0){ \makebox(\boxsize,\boxsize){$\VEC S_1$} }
        \put(8.2,0.0){ \makebox(\boxsize,\boxsize){$\VEC S_2$} }
      }

      \put(7.9,1.3){ \makebox(\boxsize,\boxsize){ $\hat{\VEC r}_{12}$ } }
      \put(7.6,1.2){ \vector(1,0){0.9}}

      \put(9.8,1.3){ \makebox(\boxsize,\boxsize){$-\VEC k$} }
      \put(10.7,1.2){ \vector(-1,0){1.2}}
    }

    \put(13.0, 1.8){
      \put(0,0){ \vector(1,0){0.9}}
      \put(0,0){ \vector(0,1){0.9}}
      \put(0.9,0.0){ \makebox(\boxsize,\boxsize){x} } 
      \put(0.0,0.8){ \makebox(\boxsize,\boxsize){y} }     
    }
  \end{picture}

  \caption{\label{fig:spin_wave_chirality}
  Spin wave chirality.
  In our \textit{ansatz}, $S^z$ is a constant of motion, therefore we represent here only the transversal components, $S^x$ and $S^y$, which change over time.
  The open circles indicate the precession sense which is fixed by the equation of motion.
  The precession phase is given by $\VEC k \cdot \VEC R_i$.
  With a spin wave, the system has two inequivalent configurations: (a) one if the wavevector is parallel to $\hat{\VEC r}_{12}$ yielding a left-handed spin wave $c_{12}=-1$ (the tilt direction is given by left-hand thumb rule); (b) another if the wavevector is antiparallel to $\hat{\VEC r}_{12}$, which results in a right-handed spin wave $c_{12}=+1$ (the tilt direction is given by right-hand thumb rule).
  }
\end{figure}
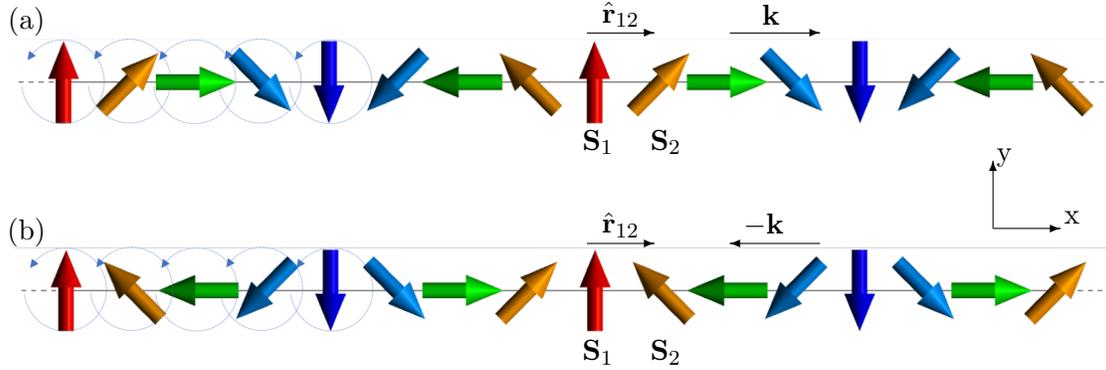

The DMI favors certain cantings between spin moments.
Let us then define a spin-wave chirality based on the canting between adjacent spins as the sign of the cross product between their projections onto the magnetization direction, and integrated over a full revolution of the precessional motion:
\begin{equation}
 c_{12}(\VEC k) 
 = \sgn\left( \int_0^\tau \hat{\VEC z} \cdot \big[\VEC S_1(\VEC k,t) \times \VEC S_2(\VEC k,t) \big] \textnormal{d}t \right) 
 =  - \sgn\left( \sin(a  \VEC k \cdot \hat{\VEC r}_{12} ) \right) \quad ,
\end{equation}
where $a$ is the lattice constant and $\hat{\VEC r}_{12}$ is a unit vector along the bond from site 1 to 2
\footnote{
Naturally, this definition depends on the choice of the spin pair.
It is important to choose a pair such that $ D^z( \VEC k) $ does not vanish for $\VEC k \parallel \hat{\VEC r}_{12}$.
}, and $\tau = 2\pi/\omega_{\VEC k}$ is the precession period.
This equation tells us that the chirality changes periodically as a function of $\VEC k$, and it is zero for $\VEC k \cdot \hat{\VEC r}_{12} = n \pi/a$ or $\VEC k \perp {\VEC r}_{12}$.
Let us take two wavevectors close to the $\Gamma$--point, one parallel and another antiparallel to $\hat{\VEC r}_{12}$, snapshots of the correspondent spin waves are shown in Figs.~\ref{fig:spin_wave_chirality}~(a) and (b), respectively.
As the DMI favors one of the two chiralities, one of the spin-wave energies is lowered while the other is raised, effectively shifting the energy minimum of the spin-wave dispersion curve out of the $\Gamma$--point in the direction of $\VEC k$ that provides the favorable chirality.
This shift is what appears as the phase $\phi$ in Eq.~\ref{eq:frequency}.

Next, let us define a more general chirality invariant that does not vary with the wavevector, which we will call the spin-wave handedness:
\begin{equation}
  \mathcal{C}_{12} = \frac{ c_{12}(\VEC k) }{ \sgn(\VEC k \cdot \hat{\VEC r}_{12}) }  \quad . 
\end{equation} 
For the spin-wave solution given by Eq.~\eqref{eq:spin_wave}, we get $\mathcal{C}_{12} = -1$.
The direction towards which the spin-wave dispersion shifts couples to the spin-wave handedness.
If the handedness were to be $+1$, instead, the shift would have been in the opposite direction.
That is the case if the spin wave were to be given by  
\begin{equation}\label{eq:spin_wave200}
\begin{split}
  \VEC S_i(\VEC k,t) =&
   \frac{ 1 }{\sqrt N} \big( 
   \cos(\VEC k \cdot \VEC R_i - \omega_\VEC k t) \hat{\VEC x}+
   \sin(\VEC k \cdot \VEC R_i - \omega_\VEC k t) \hat{\VEC y} \big)
   - S \hat{\VEC z} \quad ,
\end{split}
\end{equation}
which corresponds to a clockwise rotation and an angular momentum parallel to $\hat{\VEC z}$.
Then, we would have $\mathcal{C}_{12} = +1$.
That is, a change of handedness comes together with an inversion of the angular momentum,
and the dispersion shift due to DMI will occur in the opposite direction of that for spin waves with handedness $\mathcal{C}_{12} = -1$.
By the way, this second solution corresponds in fact to the spin waves for a ferromagnet with the magnetization along $-\hat{\VEC z}$.
This momentum-handedness coupling is imposed by the equation of motion that accepts only wave-like solutions.

As we will demonstrate in the following, this linking between angular momentum and handedness also holds for noncollinear magnetic systems, where the spatial chirality can be rather difficult to track.
Nevertheless, often these systems have excitations of very well-defined angular momentum, which will then allow us to infer their handedness and thus their response to the DMI.
This result is very powerful in allowing us to predict the effect of the DMI on the spin-wave energy and vice-versa, as we demonstrate next.

\subsection{External magnetic field and zero-net-magnetization systems}
\label{sub:external_magnetic_field_and_zero_net_magnetization}

Previously, we argued that only systems with finite net magnetization can produce a nonreciprocal spin-wave spectrum due to DMI.
Something analogous to that also happens for systems of zero net magnetization: The Dzyaloshinskii-Moriya interaction can induce chiral asymmetries in those systems too.
However, it can now only break the chirality degeneracy between rotational spin-wave modes but leaving the total spectrum reciprocally symmetric in the absence of an external magnetic field.

Let us then consider an antiferromagnet, and that the $\VEC D(\VEC k)$ aligns with the axis of the magnetic moments of the systems.
We can regard the antiferromagnet as a superposition of two coupled ferromagnetic sublattices of opposite magnetization.
Such a system has two spin-wave modes, each one with angular momentum aligned to one of the sublattice magnetizations. 
In ferromagnets, flipping the entire magnetization makes the DMI-induced asymmetry to reverse in the reciprocal space~\cite{zakeri_asymmetric_2010}.
Thus, the antiferromagnet spin waves of opposite angular momenta
are shifted in opposite directions, which effectively leaves the total spectrum of the system reciprocal.
The system becomes nonreciprocal once again under the action of an external magnetic field parallel to the alignment axis of the magnetic moments~\cite{gitgeatpong_nonreciprocal_2017,weber_non-reciprocal_2018}.
And here we have the first means through which one can reveal the asymmetry induced by DMI in systems of zero net magnetization.


\subsection{Role of spin-polarized/resolved inelastic scattering}
\label{sub:spin_resolved_inelastic_electron_scattering_role}

Now we know that DMI can induce hidden chiral asymmetry in the spin-wave spectrum in a system of zero net magnetization and that an external magnetic field can be used to reveal it.
We proceed by demonstrating that in the absence of an external magnetic field, we still can identify these asymmetries utilizing spin-polarized/resolved scattering experiments.

Often, zero-magnetization systems, such as spin spirals and antiferromagnets, host spin-wave modes that come in pairs, where the counter-partner has opposite angular momentum, and therefore, opposite handedness, e.g., two rotational modes of opposite angular momentum.
In the absence of DMI, these modes are degenerate and reciprocally symmetric, which would be the case of the two modes in an antiferromagnet.
But, as we have seen in the previous subsection, this degeneracy can be lifted by the DMI, leaving each mode nonreciprocal while the total spectrum remains reciprocal.
As we have also seen, an external magnetic field couples differently to each mode, energetically favoring one and disfavoring the other, which generates an overall nonreciprocal spectrum~\cite{gitgeatpong_nonreciprocal_2017}.

An alternative way to couple with the angular momentum of the spin waves is utilizing spin-resolved scattering experiments, such as SREELS~\cite{dos_santos_spin-resolved_2018}. 
In the example of an antiferromagnet, this would allow us to measure each mode separately by aligning the polarization of the probing particles to the precession axis of one of the spin-wave modes and measuring only the spin-flip channel.
Similarly, the same perfect mode selection can be achieved for spin-spiral systems~\cite{dos_santos_spin-resolved_2018}.
This makes of spin-polarized/resolved inelastic scattering a second means through which one can reveal the DMI-induced nonreciprocity on the spin-wave spectrum.

Next, we conjecture the conditions that rule the occurrence or not of nonreciprocal spin-wave spectra in inelastic-scattering experiments:

{(i) Nonreciprocal spectrum only occurs for spin-wave modes of finite angular momentum.}
This is a generalization of the requirement that a system needs a finite magnetization to feature a total nonreciprocal spin-wave spectrum induced by DMI in item (A).
However, this general rule applies to zero-net-magnetization systems.
As we have seen, angular momentum translates into the chiral handedness of the spin wave.
Without angular momentum, a spin wave is nonchiral and cannot manifest nonreciprocity due to DMI. 

{(ii) Only spin-flip channels may present a nonreciprocal spectrum.}
This is a direct consequence of item (D - i).
If only modes of finite angular momenta can be nonreciprocal, and usually these modes are paired to modes of opposite angular momenta, only in a spin-flip channel we can measure one disregarding the other.

{(iii) Only the component $\VEC D(\VEC k)$ parallel to the spin-wave angular momentum can influence its nonreciprocal spectrum.}

{(iv) Only a scattering experiment with the polarization of the probing particles aligned along the spin-wave angular momentum can reveal the nonreciprocity of this mode.}

\begin{figure}[t!]
\setlength{\unitlength}{1cm} 
\newcommand{\boxsize}{0.3}

  \begin{picture}(15,13)
    \put( 0, 6.5){ \includegraphics[width=15 cm,trim={0 0 0 0},clip=true]{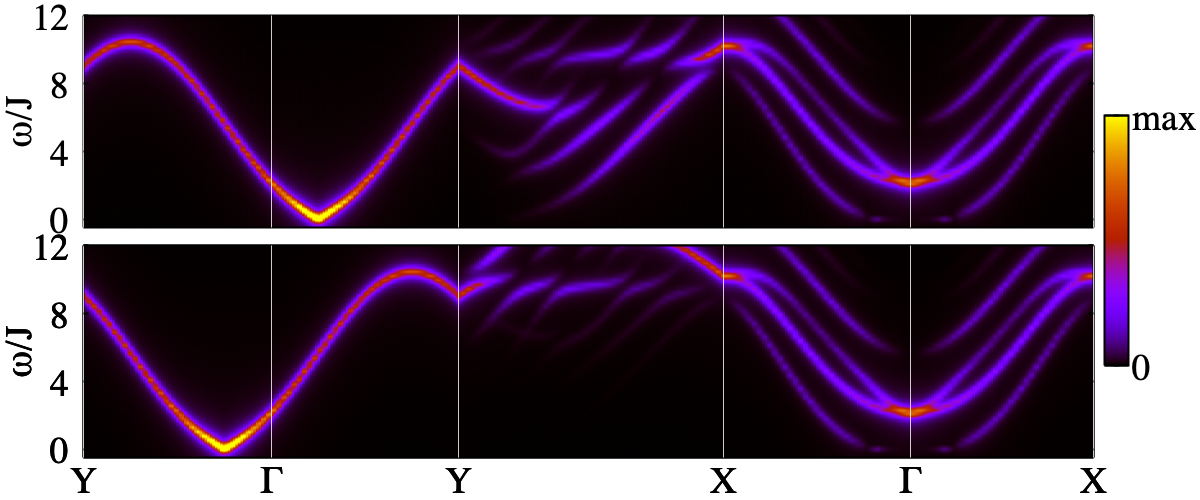} }
    \put( 0, 0.0){ \includegraphics[width=15 cm,trim={0 0 0 0},clip=true]{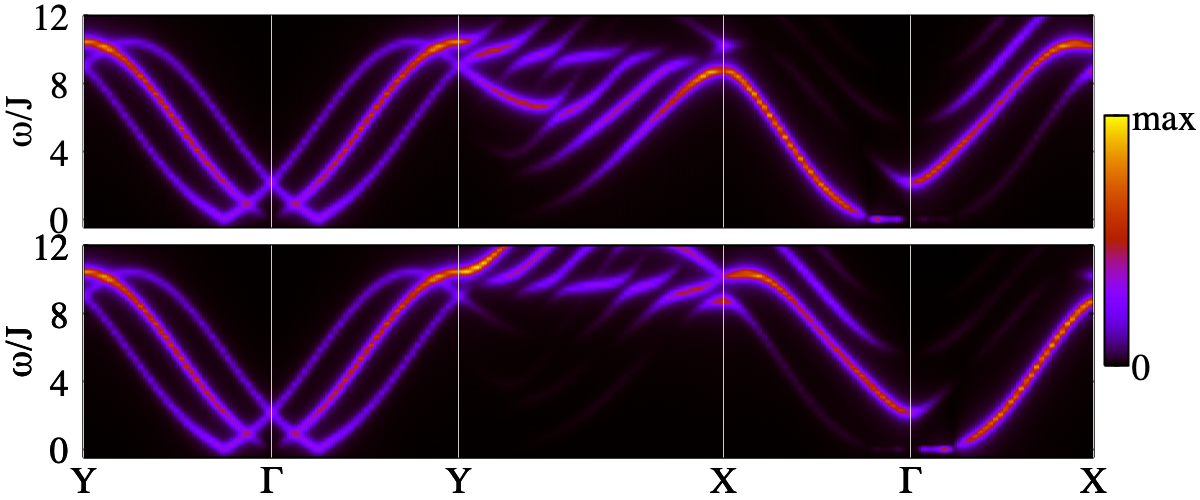} }

    \put(7.5, 10.){ \includegraphics[height=0.8 cm,trim={0 0 0 0},clip=true]{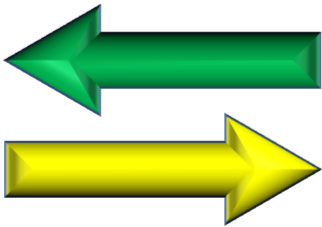} }
    \put(7.5, 7.8){ \includegraphics[height=0.8 cm,trim={0 0 0 0},clip=true]{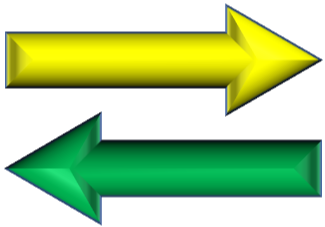} }
    \put(7.6, 3.5){ \includegraphics[width=0.8 cm,trim={0 0 0 0},clip=true]{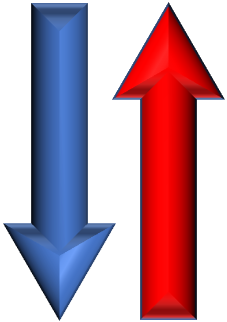} }
    \put(7.6, 0.6){ \includegraphics[width=0.8 cm,trim={0 0 0 0},clip=true]{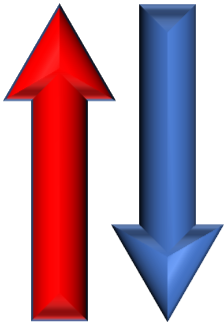} }

    \put(-0.2,12.3){ \makebox(\boxsize,\boxsize){(a)} } 
    \put(-0.2, 5.8){ \makebox(\boxsize,\boxsize){(b)} } 

    \put( 2.5,6.6){ \makebox(\boxsize,\boxsize){-\bf{Q}} } 
    \put( 3.8,6.6){ \makebox(\boxsize,\boxsize){+\bf{Q}} } 
    \put( 2.5,0.1){ \makebox(\boxsize,\boxsize){-\bf{Q}} } 
    \put( 3.8,0.1){ \makebox(\boxsize,\boxsize){+\bf{Q}} }

    \newcommand{\boxsiz}{0.3}
    \linethickness{1pt}
    \put(4.7, 7.5){
      \color{white}
      \scalebox{1}{
        \put(0,0  ){ \line(1,0){1}}
        \put(0,0.5){ \line(1,0){1}}
        \put(0,1  ){ \line(1,0){1}}

        \put(0  ,0){ \line(0,1){1}}
        \put(0.5,0){ \line(0,1){1}}
        \put(1  ,0){ \line(0,1){1}}

        \put(-0.35, 0.4){ \makebox(\boxsiz,\boxsiz){X} } 
        \put( 1.05, 0.4){ \makebox(\boxsiz,\boxsiz){X} } 
        \put( 0.40,-0.35){ \makebox(\boxsiz,\boxsiz){Y} }     
        \put( 0.40, 1.07){ \makebox(\boxsiz,\boxsiz){Y} }     
        \put(    0.5, 0.55){ \makebox(\boxsiz,\boxsiz){$\Gamma$} }     

        \put(1.55, -0.2){
          \put(0,0){ \vector(1,0){0.7}}
          \put(0,0){ \vector(0,1){0.7}}
          \put(0.7,0.0){ \makebox(\boxsiz,\boxsiz){x} } 
          \put(0.0,0.7){ \makebox(\boxsiz,\boxsiz){y} }     
        }
      }
    }
  \end{picture}

  \caption{\label{fig:nonreciprocal_model_I}
  Spin-resolved inelastic-scattering spectra for the spin spiral generated from Model I.
  (a) Shows the two spin-flip channels for polarization along $\hat{\VEC x}$, as indicated by the horizontal arrows.
  Nonreciprocity occurs in the reciprocal space where a component of $\VEC D(\VEC k)$, polarization and angular momentum align with each other.
  For Model I on path Y--$\Gamma$--Y, $\VEC D(\VEC k)$ and the angular momentum of the spin-wave modes with minima at $\VEC k = \pm \VEC Q$ are parallel to $\hat{\VEC x}$.
  (b) Shows the case for the polarization along $\hat{\VEC y}$, indicated by the vertical arrows.
  Thus, nonreciprocity is only seen in the X--$\Gamma$--X, when $\VEC D(\VEC k) \parallel \hat{\VEC y}$ that couples to the angular momentum of those spin waves.
  }
\end{figure}
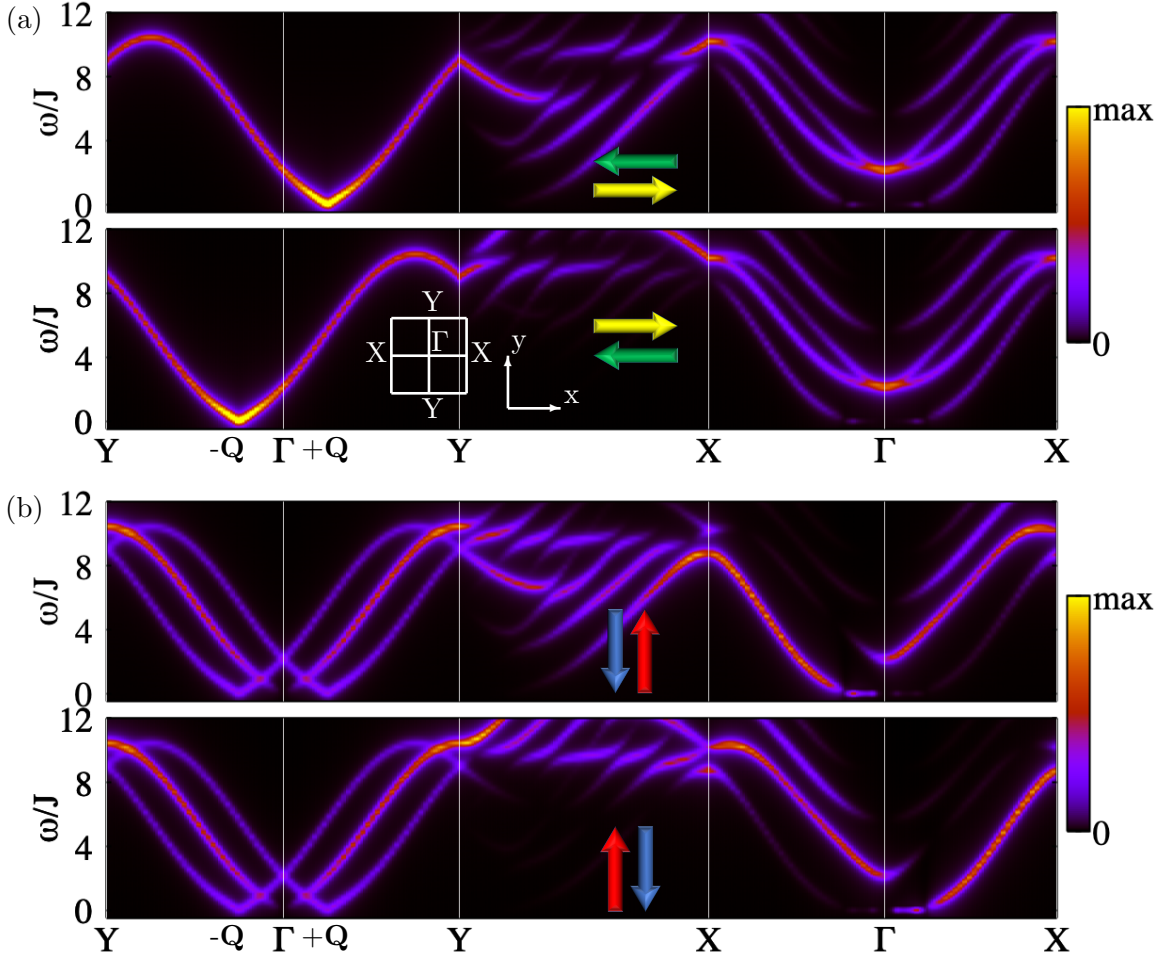

\begin{figure}[t!]
\setlength{\unitlength}{1cm} 
\newcommand{\boxsize}{0.3}

  \begin{picture}(15,13)
    \put( 0, 6.5){ \includegraphics[width=15 cm,trim={0 0 0 0},clip=true]{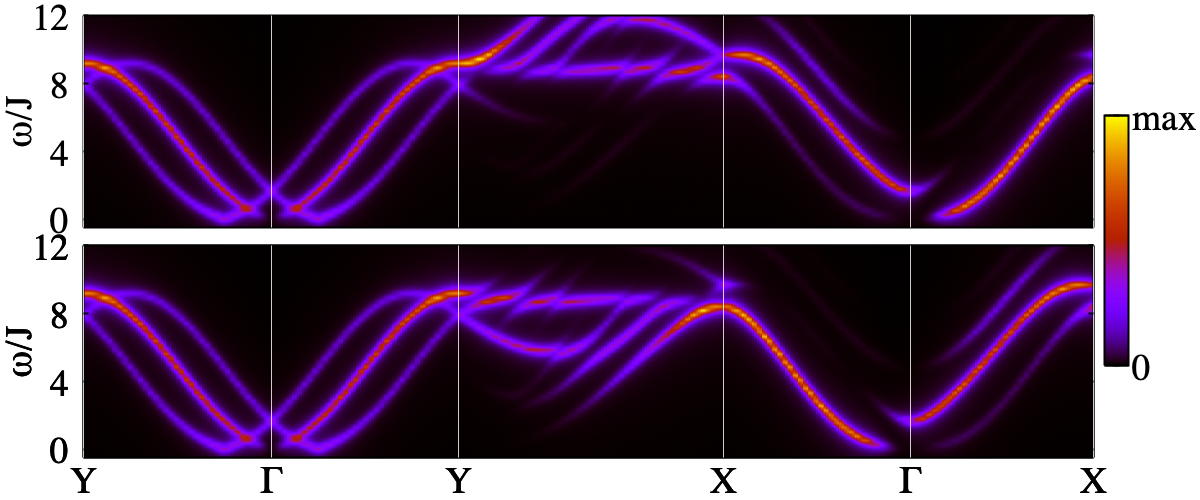} }
    \put( 0, 0.0){ \includegraphics[width=15 cm,trim={0 0 0 0},clip=true]{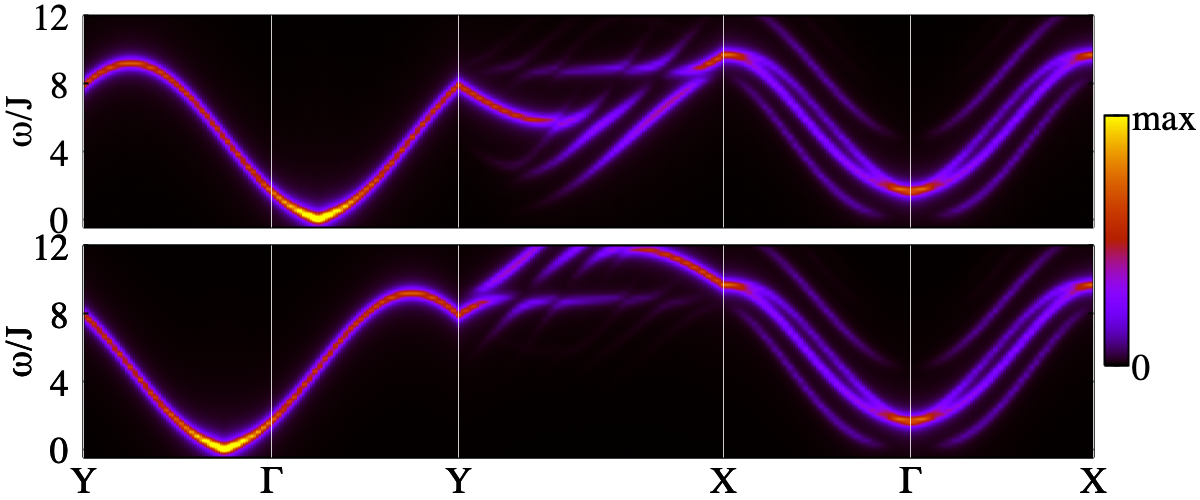} }

    \put(7.5, 10.){ \includegraphics[height=0.8 cm,trim={0 0 0 0},clip=true]{lr.png} }
    \put(7.5, 7.1){ \includegraphics[height=0.8 cm,trim={0 0 0 0},clip=true]{rl.png} }
    \put(7.6, 3.5){ \includegraphics[width=0.8 cm,trim={0 0 0 0},clip=true]{du.png} }
    \put(7.6, 1.0){ \includegraphics[width=0.8 cm,trim={0 0 0 0},clip=true]{ud.png} }

    \put(-0.2,12.3){ \makebox(\boxsize,\boxsize){(a)} } 
    \put(-0.2, 5.8){ \makebox(\boxsize,\boxsize){(b)} } 

    \put( 2.5,6.6){ \makebox(\boxsize,\boxsize){-\bf{Q}} } 
    \put( 3.8,6.6){ \makebox(\boxsize,\boxsize){+\bf{Q}} } 
    \put( 2.5,0.1){ \makebox(\boxsize,\boxsize){-\bf{Q}} } 
    \put( 3.8,0.1){ \makebox(\boxsize,\boxsize){+\bf{Q}} } 

    \newcommand{\boxsiz}{0.3}
    \linethickness{1pt}
    \put(4.7, 1.0){
      \color{white}
      \scalebox{1}{
        \put(0,0  ){ \line(1,0){1}}
        \put(0,0.5){ \line(1,0){1}}
        \put(0,1  ){ \line(1,0){1}}

        \put(0  ,0){ \line(0,1){1}}
        \put(0.5,0){ \line(0,1){1}}
        \put(1  ,0){ \line(0,1){1}}

        \put(-0.35, 0.4){ \makebox(\boxsiz,\boxsiz){X} } 
        \put( 1.05, 0.4){ \makebox(\boxsiz,\boxsiz){X} } 
        \put( 0.40,-0.35){ \makebox(\boxsiz,\boxsiz){Y} }     
        \put( 0.40, 1.07){ \makebox(\boxsiz,\boxsiz){Y} }     
        \put(    0.5, 0.55){ \makebox(\boxsiz,\boxsiz){$\Gamma$} }     

        \put(1.55, -0.2){
          \put(0,0){ \vector(1,0){0.7}}
          \put(0,0){ \vector(0,1){0.7}}
          \put(0.7,0.0){ \makebox(\boxsiz,\boxsiz){x} } 
          \put(0.0,0.7){ \makebox(\boxsiz,\boxsiz){y} }     
        }
      }
    }
  \end{picture}

  \caption{\label{fig:nonreciprocal_model_II}
  Spin-resolved inelastic-scattering spectra for the spin spiral generated from Model II.
  (a) Shows the two spin-flip channels for polarization along $\hat{\VEC x}$, as indicated by the horizontal arrows.
  Nonreciprocity occurs in the reciprocal space where a component of $\VEC D(\VEC k)$, polarization and angular momentum align with each other.
  For Model II on path X--$\Gamma$--X, $\VEC D(\VEC k)$ and the angular momentum of some spin-wave modes are parallel to $\hat{\VEC x}$.
  (b) Shows the case for the polarization along $\hat{\VEC y}$, indicated by the vertical arrows.
  Thus, nonreciprocity is only seen in the Y--$\Gamma$--Y, when $\VEC D(\VEC k) \parallel \hat{\VEC y}$ that couples to the angular momentum of the spin-wave modes whose energy minima are at $\VEC k = \pm \VEC Q$
  }
\end{figure}

Next, we demonstrate and exemplify items (iii) and (iv)  by calculating the spin-resolved spectra for the spin spirals that result from Models I and II, introduced in Sec.~\ref{sec:theoretical_framework}, with $\VEC Q \parallel \hat{\VEC y}$, shown respectively in Figs.~\ref{fig:nonreciprocal_model_I} and \ref{fig:nonreciprocal_model_II}.
Model I stabilizes a cycloidal spiral whose spins lay in the $y$--$z$ plane, see Fig.~\ref{fig:spin_configurations}~(a), while Model II leads to a helical spiral with spins lying in the $x$--$z$ plane, see Fig.~\ref{fig:spin_configurations}~(b).

Figure~\ref{fig:nonreciprocal_model_I}~(a) shows the spin-flip channels for polarization along $\hat{\VEC x}$ (represented by horizontal arrows), which present a nonreciprocal spectrum in the Y--$\Gamma$--Y path, i.e., in a reciprocal-space direction perpendicular to the polarization.
For Model I, $\VEC D(\VEC k) = - 2 D \sin( a k^y ) \hat{\VEC x}$ on this path, and therefore, it is parallel to the polarization and to the angular momentum of the spin-wave modes whose energy minima are at $\VEC k = \pm \VEC Q$.
For Fig.~\ref{fig:nonreciprocal_model_I}~(b), the polarization is set along $\hat{\VEC y}$ (represented by vertical arrows), and nonreciprocity is only seen for the X--$\Gamma$--X path, again because on this path $\VEC D(\VEC k) = 2 D \sin( a k^x ) \hat{\VEC y}$ is parallel to the polarization and the angular momentum of some spin-wave modes.
Naturally, a polarization along $z$ will not feature any nonreciprocity, because the DMI model has no component along that direction.

For Model II, the Fourier transformed DMI vector on path Y--$\Gamma$--Y is $\VEC D(\VEC k) = 2 D \sin( a k^y ) \hat{\VEC y}$, and along X--$\Gamma$--X it is $\VEC D(\VEC k) = 2 D \sin( a k^x ) \hat{\VEC x}$.
Thus, in contrast to Model I, we will observe the nonreciprocity on the reciprocal-space direction parallel to the polarization, see Fig.~\ref{fig:nonreciprocal_model_II}~(a) and (b), where the polarization is along $\hat{\VEC x}$ and $\hat{\VEC y}$, respectively.
In both Models, the spin-spiral wavevector points along the same direction, but the direction where the nonreciprocity occurs changes from one model to the other, which shows that the direction of the spiral wavevector has no role on the nonreciprocity.

\begin{figure}[t!]
\setlength{\unitlength}{1cm} 
\newcommand{\boxsize}{0.3}

  \begin{picture}(15,6)
    \put(  0, 0.0){ \includegraphics[width=15 cm,trim={0 0 0 0},clip=true]{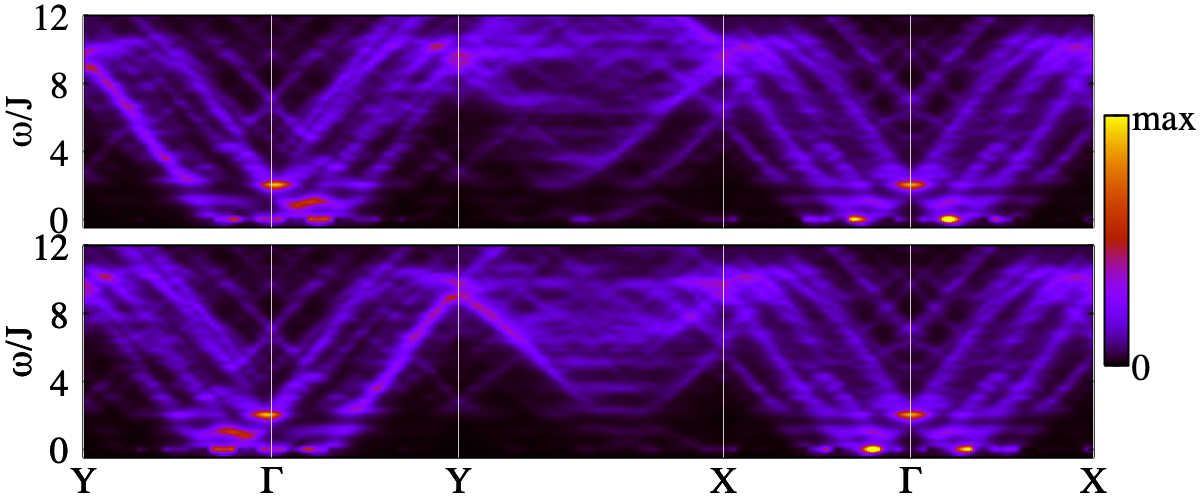} }

    \put(6.8, 3.5){ \includegraphics[height=0.8 cm,trim={0 0 0 0},clip=true]{lr.png} }
    \put(6.8, 0.6){ \includegraphics[height=0.8 cm,trim={0 0 0 0},clip=true]{rl.png} }

    \put(-0.2, 5.8){ \makebox(\boxsize,\boxsize){(a)} } 
    \put(-0.2, 2.9){ \makebox(\boxsize,\boxsize){(b)} } 
  \end{picture}

  \caption{\label{fig:nonreciprocal_model_I_SK}
  Spin-resolved inelastic-scattering spectra for a skyrmion lattice generated by Model I added with an out-of-plane external magnetic field.
  Panel (a) represents the left-right spin-flip channel, and (b) shows the right-left one, as indicated by the horizontal arrows.
  The beam polarization is along $\hat{\VEC x}$.
  The spectra resemble a continuum of excitations rather than well-defined dispersing lines.
  Nevertheless, the nonreciprocity is visible (along the Y--$\Gamma$--Y path for this beam polarization), and their occurrence conditions match those for the spin spiral established by the same DMI model in the absence of the external field, see also Fig.~\ref{fig:nonreciprocal_model_I}~(a). 
  }
\end{figure}

For more complex systems with lower symmetries, such as skyrmion lattices, the spectrum of each spin-wave mode is not well-defined throughout the reciprocal space in inelastic-scattering experiments.
The spectra are closer to a continuum of excitations instead of the well-separated branches seen for the spin-spiral configurations, see Fig.~\ref{fig:nonreciprocal_model_I_SK}, whereupon adding an out-of-plane external magnetic field to Model I could stabilize a skyrmion lattice in an $8\times8$ atoms unit cell, see also Fig.~\ref{fig:spin_configurations}~(c).
Naturally, it is also hard to identify the direction of the angular momentum of the underlying spin wave corresponding to each high-intensity region of the spectrum.
Nevertheless, the nonreciprocity is still present and measurable.
In Fig.~\ref{fig:nonreciprocal_model_I_SK}, we observe a nonreciprocity on the same path, Y--$\Gamma$--Y, as seen for the spin spiral established in the absence of the external magnetic field, see Fig.~\ref{fig:nonreciprocal_model_I}~(a), for the same polarization along $\hat{\VEC x}$.
Even though the two systems look rather different from each other, the reciprocity on their spectra occurs under the same condition because they share the same DMI structure.

As we have demonstrated, only spin-flip channels can present a nonreciprocal spectrum.
However, not always a spin-resolved inelastic-scattering experiment is available, as is currently the case of electron scattering setup to study spin waves.
A more easily accessible experiment is the spin-polarized setups, where a source of spin-polarized particles is used to scatter from the magnetic material and the spin of the scattered particle is not measured.
The resulting spectrum is equivalent to the addition of a spin-flip and a non-spin-slip channel, e.g., down-up plus down-down.
While the latter cannot be nonreciprocal, the first can and so is their sum.

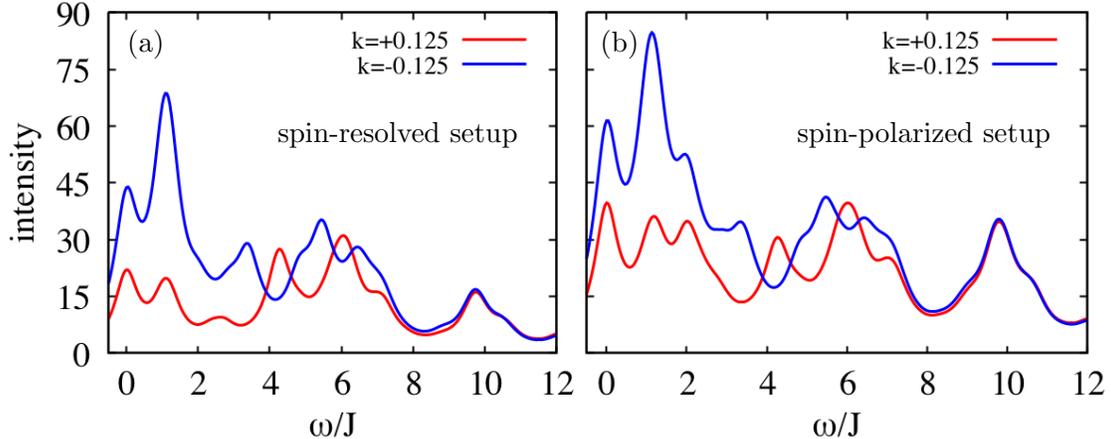
\begin{figure}[t!]
\setlength{\unitlength}{1cm} 
\newcommand{\boxsize}{0.3}

  \begin{picture}(15,6)
    \put(  0, 0.0){ \includegraphics[width=15 cm,trim={0 0 0 0},clip=true]{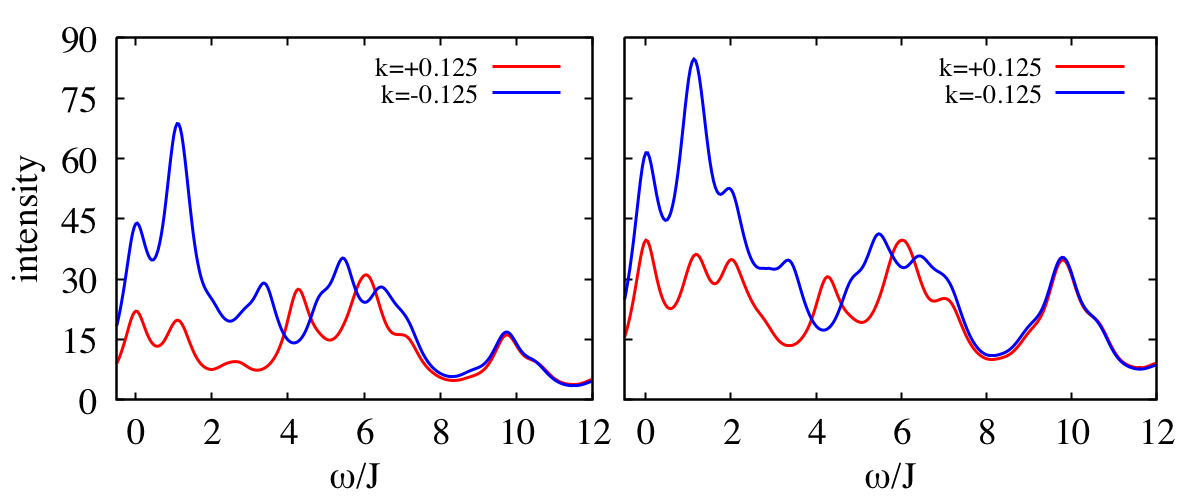} }

    \put( 3.8, 4.0){ \makebox(3,\boxsize){spin-resolved setup} }
    \put(10.8, 4.0){ \makebox(3,\boxsize){spin-polarized setup} }

    \put( 1.8, 5.2){ \makebox(\boxsize,\boxsize){(a)} }
    \put( 8.1, 5.2){ \makebox(\boxsize,\boxsize){(b)} }
  \end{picture}

  \caption{\label{fig:constq_curves}
  Constant-wavevector inelastic-scattering spectra for a skyrmion lattice generated by Model I added with an out-of-plane external magnetic field.
  The spectra were calculated for two opposite wavevector $\VEC k = \pm  \frac{2\pi}{a} k\,\hat{\VEC y}$.
  The beam polarization is along $\hat{\VEC x}$.
  (a) shows the spin-resolved setup, where only one spin-flip channel is taken (left-right scattering channel).
  (b) presents the spin-polarized setup, which results from adding a spin-flip and a non-spin-flip (left-right + left-left scattering channels).
  In both cases, we can observe that the inelastic signal at $-\VEC k$ is distinct and predominantly higher than at $\VEC k$, therefore, it is nonreciprocal.
  The multiple peaks correspond to the various spin-wave modes of the skyrmion lattice, in contrast to the expected single peak for a ferromagnetic phase and the three modes of a spin spiral.
  }
\end{figure}

Figure~\ref{fig:constq_curves} represents constant wave-vector spectra, which are the typical measurements done in inelastic electron and neutron scattering experiments.
The wavevector of the spin excitations are fixed by controlling the ratio between the incident and scattering angles, and the intensity corresponds to the number of probing particles that have transferred a given amount of energy to the excitations in an interval of time.
We calculated the spectra for wavevectors opposite to each other in the reciprocal space, $\VEC k = \pm 2\pi k \hat{\VEC y}$, and the polarization was set along the $\hat{\VEC x}$ direction, which aligns with $\VEC D(\VEC k)$.
Figure~\ref{fig:constq_curves}~(a) shows the results for a spin-resolved setup (which corresponds to a vertical line of the spectrum shown in Fig.~\ref{fig:nonreciprocal_model_I_SK}~(a)), while Fig.~\ref{fig:constq_curves}~(b) presents the spin-polarized spectrum.
In the low-energy region, we clearly observe for both setups, spin-resolved or spin-polarized, a difference in the scattering intensity.
For higher energies, some peaks vanish and others appear when comparing the spectra for the two opposite wavevectors.

It is the DMI directional sense that determines which scattering intensity will be higher, at $+\VEC k$ or $-\VEC k$.
Upon reversing the DMI, the spectra would be swapped in Fig.~\ref{fig:constq_curves}.
This implies that such an experiment measures the DMI sense.

Let us take Model I with an out-of-plane magnetic field which stabilized a skyrmion lattice, and now reverse the chirality of the DMI along one direction only, making $D^x \rightarrow -D^x$.
This modified model then stabilizes an antiskyrmion lattice.
Because the skyrmion and antiskyrmion systems translate into the other only by a mirror reflection operation, their total spin-wave spectra, which are reciprocal, do not differ.
However, as we have shown, the scattering rate can depend directly on the DMI orientation, and we should be able in this case to identify it.


\subsection{Dzyaloshinskii-Moriya interaction and spin-wave angular momentum}
\label{sub:dz_and_spin_wave_angular_momentum}

We saw that the nonreciprocity is seen when the probing-beam polarization, the DMI vector in reciprocal space $\VEC D(\VEC k)$, and the spin wave's angular momentum align.
It is easy to see that the polarization couples to the angular momentum, however, how does the angular momentum couple to the DMI?
Is the angular momentum, which is the property that allows the nonreciprocal inelastic measurement, given by the spin structure or by the Dzyaloshinskii-Moriya interactions?
The answer is that both, spin configuration and the DMI set the angular momentum of the spin waves.

\begin{figure}[t!]
\setlength{\unitlength}{1cm} 
\newcommand{\boxsize}{0.3}

  \begin{picture}(15,6)
    \put(  0, 0.0){ \includegraphics[width=15 cm,trim={0 0 0 0},clip=true]{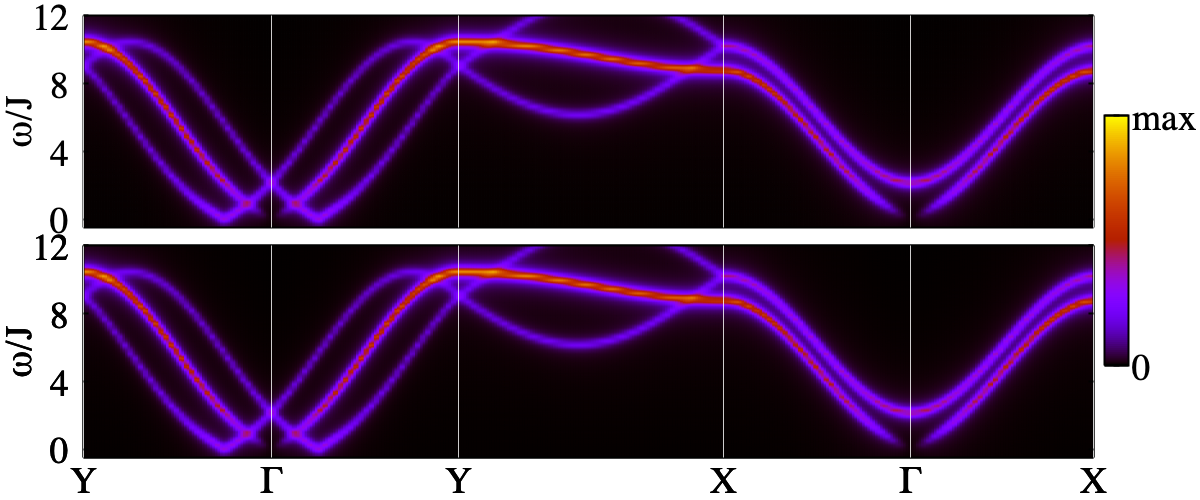} }

    \put(7.0, 3.5){ \includegraphics[width=0.8 cm,trim={0 0 0 0},clip=true]{du.png} }
    \put(7.0, 0.6){ \includegraphics[width=0.8 cm,trim={0 0 0 0},clip=true]{ud.png} }

    \put( 2.5,0.1){ \makebox(\boxsize,\boxsize){-\bf{Q}} } 
    \put( 3.8,0.1){ \makebox(\boxsize,\boxsize){+\bf{Q}} } 
  \end{picture}

  \caption{\label{fig:nonreciprocal_model_I_Dy0}
  Spin-resolved inelastic-scattering spectra for the spin spiral generated by Model I with $D^y=0$.
  Note that the spiral itself is stabilized by $D_x$.
  The polarization is set along $\hat{\VEC y}$, as indicated by the vertical arrows.
  The two spin-flip channels are degenerate and reciprocal because $\VEC D(\VEC k)$ has no component along the polarization to induce angular momentum of the spin-wave modes along that direction.
  Restoring the $D^y$ of the original Model I, a nonreciprocity occurs on the X--$\Gamma$--X while the ground-state spin configuration is not affected, proving that the DMI can directly induce the nonreciprocity of spin waves.
  }
\end{figure}

Let us consider Model I with $D^y$ set to zero.
The same cycloidal spin spiral with $\VEC Q \parallel \hat{\VEC y}$ is still the ground state.
Previously, we have seen on Fig.~\ref{fig:nonreciprocal_model_I} (b) that spin-resolved inelastic-scattering spectra for polarization along $\hat{\VEC y}$ featured nonreciprocity in the X--$\Gamma$--X path because there $\VEC D(\VEC k)$ was parallel to $\hat{\VEC y}$.
Now, once $\VEC D(\VEC k) = 0$ on that same path, the spectrum becomes reciprocal, see Fig.~\ref{fig:nonreciprocal_model_I_Dy0}.
This proves that the nonreciprocity is not only induced by the spin structure but also directly by the DMI.
Similarly, one observes that a spin spiral stabilized by exchange interaction frustration, without involving DMI, can also feature nonreciprocity as if the DMI that could favor that structure were there~\cite{cheon_nonreciprocal_2018}.

We have seen that the DMI only influences the dispersion and the inelastic spectra of spin-wave modes whose angular momenta have a finite projection on $\VEC D(\VEC k)$.
An antiferromagnet hosts two counter-rotating spin-wave modes that precess in the plane perpendicular to the axis of the magnetic moments.
That is why we discussed in item (iii) an example where $\VEC D(\VEC k)$ is parallel to this axis, which guarantees that the DMI would maximally influence the spin-wave modes.
However, for a general noncollinear magnetic structure, the angular momenta of the spin waves are not obvious, and thus, only knowing the DMI structure will not be enough to predict the occurrence of the asymmetries.
Our observations have shown, however, that spin structures that are energetically favored by a given component of $\VEC D(\VEC k)$ will host spin waves whose angular momenta are along this same DMI component, i.e., they will also induce nonreciprocity to the system.
We can exemplify this by taking the Models I and II again, and the spin spirals that each one favors as the ground state.
For Model I, the cycloidal spiral with spins lying in the $yz$-plane is stabilized by $D^x$, and so the angular momenta of the $+\VEC Q$ and $-\VEC Q$ modes are along $\hat{\VEC x}$.
Meanwhile, the helical spiral of Model II is stabilized by $D^y$, and its $\pm \VEC Q$ modes have angular momenta along $\hat{\VEC y}$.



\section{Conclusions} 

In this article, we contributed to the problem of mapping the Dzya\-lo\-shin\-skii-Moriya interaction in systems of complex magnetic structures.
We did that by studying the effect of the DMI on the dynamics of spin waves.
We made an important connection between the angular momentum and the chiral handedness of a spin-wave mode.
Effectively, this allows us to predict when a given spin-wave mode energy and scattering rate is affected by the DMI.

We saw that the DMI can induce nonreciprocity in the spin waves.
We concluded that only systems of finite magnetization can have a total spin-wave spectrum that is nonreciprocal.
Nevertheless, nonreciprocity can also occur for individual spin-wave modes in systems with zero-net-magnetization and noncollinear spin textures, while the total spectrum remains reciprocal.

We showed that an external magnetic field and spin-resolved energy loss spectroscopy (SREELS), proposed in Ref.~\cite{dos_santos_spin-resolved_2018}, can help to reveal the nonreciprocity of individual modes.
We saw that only a spin-flip scattering spectrum can present nonreciprocity and that a nonreciprocal spectrum is expected when a component of $\VEC D(\VEC k)$ is parallel to the angular momentum and the polarization of the probing electrons.
As we can control the polarization of the probe beam, and the spin-resolved measurements can also determine the angular momentum of the spin waves, ultimately we can determine the DMI chirality even for zero-net-magnetization systems.
This achievement is in contrast to previous expectations found on the literature~\cite{dmitrienko_measuring_2014}, where other authors resorted to controlling the phase and amplitude of the probing beam to be able to determine the DMI chirality.

For the case of a skyrmion lattice, despite having a finite out-of-plane net magnetization, no component of the DMI projects along that net magnetization (which is in-plane), guaranteeing that the total spin-wave spectrum is reciprocal.
Nevertheless, the scattering rate still can have nonreciprocity induced by the DMI.
This allowed us to detect a change in the chirality of the DMI along different directions, which permits us, for instance, to infer the existence of antiskyrmions instead of skyrmions~\cite{hoffmann_antiskyrmions_2017}. 

Finally, we learned that the Dzyaloshinskii-Moriya interaction can influence the angular momentum of the spin waves directly and indirectly.
In general, the DMI favors the formation of spin structures that naturally hosts spin waves whose precession axis aligns with the DMI.
That is, the spin-wave angular momenta tend to be along $\VEC D(\VEC k)$ that favored the spin configuration in the first place.
However, even those components of $\VEC D(\VEC k)$ that do not contribute to the energy of the ground state can directly influence the dynamics of spin waves, in particular of their angular momentum and thus their scattering rate.

\begin{acknowledgments}

This work is supported by the Brazilian funding agency CAPES under Project No. 13703/13-7 and the European Research Council (ERC) under the European Union's Horizon 2020 research and innovation programme (ERC-consolidator Grant No. 681405-DYNASORE). We gratefully acknowledge the computing time granted by JARA-HPC on the supercomputer JURECA at Forschungszentrum Jülich and by RWTH Aachen University.
\end{acknowledgments}

\bibliography{paper_nonreciprocity}

\begin{appendix}
\section{On the chiral asymmetry of spin waves}
\label{Apx:on_the_chiral_asymmetry_of_spin_waves}

\subsection{Spin-wave chirality in ferromagnets}
\label{Apxsec:appendix_spin_wave_chirality_in_ferromagnets}

The only contribution in the Hamiltonian that can be sensitive to the chirality of a spin wave (see Sec.~\ref{sub:nonreciprocal_spin_wave_spectrum}) is that of the Dzyaloshinskii-Moriya interaction.
It goes with the cross product of two spin moments at different sites.
If we consider the \textit{ansatz} for a spin-wave snapshot given by Eq.~\eqref{eq:spin_wave_snapshot}, we get:
\begin{equation}
\begin{split}
  \VEC{S}_i \times \VEC{S}_j = & 
   \left[ 
  (S_i^2 S_j^0 - S_i^0 S_j^2) \VEC n^1 +
  (S_i^0 S_j^1 - S_i^1 S_j^0) \VEC n^2 +
  (S_i^1 S_j^2 - S_i^2 S_j^1) \VEC n^0 
  \right]  \\
  = & \left[ 
  c \sin\theta \cos\theta ( \sin\phi_i  -  \sin\phi_j ) \VEC n^1   
     + \cos\theta \sin\theta ( \cos\phi_j - \cos\phi_i  ) \VEC n^2 \right. \\
    &+ \left. c \sin^2\theta \sin(\VEC k \cdot (\VEC R_j - \VEC R_i))  \VEC n^0 
  \right] \quad ,
\end{split}
\end{equation}
and therefore, two terms depend on the chirality constant $c$.
However, evaluating the sum over all lattice points required by the hamiltonian Eq.~\eqref{eq:HeisenbergHamiltonian}, the first term vanishes:
\begin{equation}
\begin{split}
\sum_{ij}D^1_{ij}( \sin\phi_i  -  \sin\phi_j ) = 0 \\
\sum_{ij} 2 D^1_{ij} \sin\phi_i  = 0 \quad ,
\end{split}
\end{equation}
because $D^1_{ij} = - D^1_{ji}$.

Thus, the only term that depends on the spin-wave chirality in the energy, obtained by substituting the spin-wave equation of Eq.~\eqref{eq:spin_wave_snapshot} into the Hamiltonian in Eq.~\eqref{eq:HeisenbergHamiltonian}, has the form
\begin{equation}
\begin{split}
E(\VEC k, c) 
  = & - \frac{1}{2} c \sin^2\theta \sum_{ij}  \sin(\VEC k \cdot (\VEC R_j - \VEC R_i) \VEC D_{ij} \cdot\VEC n^0 
  = - \frac{1}{2} c \sin^2\theta N \VEC n^0 \cdot  \VEC D(\VEC k)  \quad ,
\end{split}
\end{equation}
where
\begin{equation}
  \VEC D(\VEC k) = \sum_{j}  \sin(\VEC k \cdot \VEC R_{ij}) \VEC D_{ij} \quad .
\end{equation}
We can notice that only the $\VEC D(\VEC k)$ component along the magnetization contributes to the chirality.
This result matches the conclusion of L. Udvardi and L. Szunyogh~\cite{udvardi_chiral_2009}.

\subsection{Spin waves in a classical approach} 
\label{sec:spin_waves_in_a_classical_approach}

In this section, we solve the equation of motion for every spin in a ferromagnet to understand the dynamics of its spin waves and the corresponding local spin precession.

\subsubsection{Effective field} 
\label{sub:effective_field}

Considering the magnetic moments of a ferromagnet as classical vectors, their dynamics are governed by the phenomenological equation of motion given by Eq.~\eqref{eq:eq_motion}.
Solving this equation simultaneously for all sites provides spin-wave solutions.
First, we need to determine the effective field, given by Eq.~\eqref{eq:eff_field}, which for the Hamiltonian of Eq.~\eqref{eq:HeisenbergHamiltonian} reads
\begin{equation}
\begin{split}
  B^\text{eff}_i =& - \partdv{H}{\VEC S_i} \\
  =& \frac{1}{2} \partdv{}{\VEC S_i} \sum_{kj} \left(  J_{kj} \VEC S_k \cdot \VEC S_j + \VEC D_{kj} \cdot (\VEC S_k \times \VEC S_j) \right) + \VEC B\\
  =& \frac{1}{2} \partdv{}{\VEC S_i} \left[ 
  \sum_{j} \left(  J_{ij} \VEC S_i \cdot \VEC S_j +  \VEC S_i \cdot (\VEC S_j \times \VEC D_{ij})  \right) \right .+
  \sum_{k} \left(  J_{ki} \VEC S_k \cdot \VEC S_i +  \VEC S_i \cdot (\VEC D_{ki} \times \VEC S_k)   \right)  \\
   & +  \left.
  \sum_{k\neq i,j\neq i} \left(  J_{kj} \VEC S_k \cdot \VEC S_j + \VEC D_{kj} \cdot \VEC S_k \times \VEC S_j \right)
   \right] + \VEC B  \\
  =& \frac{1}{2}\sum_{j} \left(  J_{ij} \VEC S_j +   \VEC S_j \times \VEC D_{ij}   \right)  
   + \frac{1}{2}\sum_{j} \left(  J_{ji} \VEC S_j +   \VEC D_{ji} \times \VEC S_j   \right) + \VEC B \\
  =& \sum_{j} \left(  J_{ij} \VEC S_j +   \VEC S_j \times \VEC D_{ij}   \right) + \VEC B \quad . 
\end{split}
\end{equation}
In calculating the derivative of the Hamiltonian, we did not have to take care of terms with $k=j=i$ because $J_{ij}$ and $\VEC D_{ij}$ are zero.
Also, we made use of the cyclic permutation of the scalar triple product: $\VEC a \cdot (\VEC b \times \VEC c ) = \VEC c \cdot (\VEC a \times \VEC b) = \VEC b \cdot (\VEC c \times \VEC a) $;
and we swapped the interaction parameters index respecting their symmetries: $J_{ij}=J_{ji}$ and $\VEC D_{ij} = - \VEC D_{ji}$.

\subsubsection{Equation of motion} 
\label{sub:equation_of_motion}

Thus, the equation of motion in Eq.~\eqref{eq:eq_motion} reads
\begin{equation}
\begin{split}
  \dv{\VEC S_i}{t} 
  =& - \sum_{j} \left(  J_{ij} \VEC S_i \times \VEC S_j +   \VEC S_i \times (\VEC S_j \times \VEC D_{ij} )  \right) -\VEC S_i \times \VEC B\\
  =& - \sum_{j} \left(  J_{ij} \VEC S_i \times \VEC S_j +   \VEC S_j (\VEC S_i \cdot \VEC D_{ij} ) -   \VEC D_{ij}  (\VEC S_i  \cdot \VEC S_j)  \right) -\VEC S_i \times \VEC B\\
  =
   & - \sum_j  \left[    J_{ij} \left( S_i^y S_j^z - S_i^z S_j^y \right) + S_j^x (\VEC S_i \cdot \VEC D_{ij} ) -   D_{ij}^x  (\VEC S_i  \cdot \VEC S_j)  \right] \hat{\VEC x} - ( S_i^y B^z - S_i^z B^y) \hat{\VEC x}  \\
   & - \sum_j  \left[    J_{ij} \left( S_i^z S_j^x - S_i^x S_j^z \right) + S_j^y (\VEC S_i \cdot \VEC D_{ij} ) -   D_{ij}^y  (\VEC S_i  \cdot \VEC S_j)  \right] \hat{\VEC y} - ( S_i^z B^x - S_i^x B^z) \hat{\VEC y}   \\
   & - \sum_j  \left[    J_{ij} \left( S_i^y S_j^x - S_i^x S_j^y \right) + S_j^z (\VEC S_i \cdot \VEC D_{ij} ) -   D_{ij}^z  (\VEC S_i  \cdot \VEC S_j)  \right] \hat{\VEC z} - ( S_i^x B^y - S_i^y B^x) \hat{\VEC z}  \quad .
\end{split}
\end{equation}
Let us assume a magnetic field of magnitude $B$ along the $z$ direction and that the motion of each spin is of a small amplitude around the equilibrium axis.
This implies that we consider that $S_i^x, S_i^y \ll 1$, and in first-order approximation we disregard all products between them and take $S^z_i \sim S$.
Thus, the above equation becomes
\begin{equation}\label{eq:eq_motion2}
\begin{split}
  \dv{\VEC S_i}{t} 
  =
   & \Big( - S \sum_j  \left[   J_{ij} \left( S_i^y - S_j^y \right) + D_{ij}^z S_j^x  -  S D_{ij}^x    \right] - B S_i^y  \Big ) \hat{\VEC x}  \\
   & \Big( - S \sum_j  \left[   J_{ij} \left( S_j^x - S_i^x \right) + D_{ij}^z S_j^y  -  S D_{ij}^y    \right] + B S_i^x  \Big ) \hat{\VEC y} \\
   & - S \sum_j  \left[   D_{ij}^x S_i^x  +  D_{ij}^y S_i^y   \right] \hat{\VEC z} \\
  =
   & \Big ( -S \sum_j  \left[ J_{ij} \left( S_i^y - S_j^y \right) + D_{ij}^z S_j^x   \right] - B S_i^y  \Big ) \hat{\VEC x} \\
   & \Big ( -S \sum_j  \left[ J_{ij} \left( S_j^x - S_i^x \right) + D_{ij}^z S_j^y   \right] + B S_i^x  \Big ) \hat{\VEC y}  \quad ,
\end{split}
\end{equation}
because $\sum_j D_{ij}^{x,y} = 0$ when the summation is over a Bravais lattice due to the antisymmetry of the DMI.
We can see that, within the linear approximation, only the component of DMI along the magnetization matters.

This is a vectorial equation, which represents two equations: one for the $x$ and one for the $y$ components of the spin moment.
Here note that the dynamics of one of the components depends on that of the other, therefore, we have a set of two coupled equations.
Then, let us consider the following transformation:
\begin{equation} \label{eq:circ_components}
\begin{split}
  S_i^+ = S_i^x + \iu S_i^y \quad \text{and}& \quad S_i^- = S_i^x - \iu S_i^y \\
  S_i^x = \frac{1}{2} \left( S_i^+ + S_i^- \right) \quad \text{and}& \quad S_i^y = \frac{1}{2\iu} \left(S_i^+  - S_i^-\right) \quad ,
\end{split}
\end{equation}
which define the circular components of the spin moments.
Applying that to Eq.~\eqref{eq:eq_motion2}, we find
\begin{equation}\label{eq:eq_motion3}
\begin{split}
 \iu \dv{ (S_i^+ + S_i^-) }{t} 
  =& S \sum_j  \left[ J_{ij} \left(-S_i^+ + S_i^- + S_j^+ - S_j^- \right) - \iu D_{ij}^z (S_j^+ + S_j^-)   \right] - B (S_i^+ -S_i^-) \quad   \\
 \iu \dv{ (S_i^+ - S_i^-) }{t}
  =& S \sum_j  \left[ J_{ij} \left(-S_i^+ - S_i^- + S_j^+ + S_j^- \right) - \iu D_{ij}^z (S_j^+ - S_j^-)   \right] - B (S_i^+ +S_i^-) \quad . \\
\end{split}
\end{equation}
Combining these two equations, we get
\begin{equation}
\begin{split}
-\iu \dv{ S_i^+}{t} 
  =& S \sum_j  \left[ J_{ij} \left(S_i^+ - S_j^+ \right) + \iu D_{ij}^z S_j^+   \right] + B S^+_i \quad \\
 \iu \dv{ S_i^- }{t}
  =& S \sum_j  \left[ J_{ij} \left(S_i^- - S_j^- \right) - \iu D_{ij}^z S_j^-   \right] + B S^-_i \quad , \\
\end{split}
\end{equation}
which can be simplified by introducing the following definition
\begin{equation}
  J_{ij}^\pm = J_{ij} \pm \iu D_{ij}^z \quad ,
\end{equation}
that allow us to write
\begin{equation}\label{eq:eq_motion4}
\begin{split}
-\iu \dv{ S_i^+}{t} 
  =& S \sum_j  \left[  J_{ij} S_i^+ - J_{ij}^- S_j^+   \right]+ B S^+_i         \\
 \iu \dv{ S_i^- }{t}
  =& S \sum_j  \left[  J_{ij} S_i^- - J_{ij}^+ S_j^-  \right] + B S^-_i \quad , \\
\end{split}
\end{equation}
defining two \emph{decoupled} equations of motion.

\subsubsection{Fourier transformation} 
\label{sub:fourier_transformation}
  
The dynamics of a given site depends on what is happening to all sites connected to it via the exchange interaction.
However, if the system has translational symmetry, we can Fourier transform these equations defining
\begin{equation}\label{eq:Four_transf}
\begin{split}
  S_{\VEC k}^\pm = \frac{1}{\sqrt{N}} \sum_i e^{- \iu \VEC k \cdot \VEC R_i} S_i^\pm \quad , \quad
  S_{i     }^\pm = \frac{1}{\sqrt{N}} \sum_{\VEC k} e^{  \iu \VEC k \cdot \VEC R_i} S_{\VEC k}^\pm \quad . \\
\end{split}
\end{equation}
Then, by left multiplying Eq.~\eqref{eq:eq_motion4} with $ \frac{1}{\sqrt{N}}\sum_i e^{- \iu \VEC k \cdot \VEC R_i}$, one gets
\begin{equation}\label{eq:eq_motion5}
\begin{split}
  -\iu \dv{ S_{\VEC k}^+(t) }{t} =& \frac{S}{\sqrt{N}}  \left(\sum_i e^{-\iu \VEC k \cdot \VEC R_i } S_i^+(t) \sum_j  J_{ij}  - \sum_j e^{- \iu \VEC k \cdot \VEC R_j } S_j^+(t) \sum_i e^{+ \iu \VEC k \cdot (\VEC R_j - \VEC R_i) } J_{ij}^-  \right) + B S^+_{\VEC k}(t)   \\
   \iu \dv{ S_{\VEC k}^-(t) }{t} =& \frac{S}{\sqrt{N}}  \left(\sum_i e^{-\iu \VEC k \cdot \VEC R_i } S_i^-(t) \sum_j  J_{ij}  - \sum_j e^{- \iu \VEC k \cdot \VEC R_j } S_j^-(t) \sum_i e^{+ \iu \VEC k \cdot (\VEC R_j - \VEC R_i) } J_{ij}^+  \right) + B S^-_{\VEC k}(t) \quad ,  \\
\end{split}
\end{equation}
where we multiplied the second term of the r.h.s by 1 in the form of $ e^{- \iu \VEC k \cdot \VEC R_j } e^{ \iu \VEC k \cdot \VEC R_j}$.
We then obtain
\begin{equation}\label{eq:eq_motion6}
\begin{split}
  -\iu \dv{ S_{\VEC k}^+(t) }{t} =& \Big( S \left( J_{\VEC 0} - J_{\VEC k}^- \right) + B \Big) S_{\VEC k}^+(t)   \\
   \iu \dv{ S_{\VEC k}^-(t) }{t} =& \Big( S \left( J_{\VEC 0} - J_{\VEC k}^+ \right) + B \Big) S_{\VEC k}^-(t) \quad ,  \\
\end{split}
\end{equation}
where the Fourier transformed interactions is defined as
\begin{equation}\label{eq:Four_J}
\begin{split}
 J_{\VEC k}^\pm 
 = \sum_i e^{ \iu \VEC k \cdot \VEC R_{ij} } J_{ij}^\pm \quad ,
\end{split}
\end{equation}
which assumes a translational symmetry, such that $J_{ij}^\pm $ only depends on the difference $\VEC R_{ij} = \VEC R_j - \VEC R_i$.
Note as well, that $J_{\VEC 0}^\pm = \sum_i ( J_{ij} \pm \iu D^z_{ij}) = \sum_i J_{ij} = J_\VEC 0$, again because of the DMI antisymmetry.
Next follows some useful properties of the interactions in the reciprocal space:
\begin{equation}\label{eq:Four_J2}
\begin{split}
J_{-\VEC k}^\pm &
=   \sum_i e^{-\iu \VEC k \cdot \VEC R_{ij} } J_{ij}^\pm 
=   \sum_i e^{ \iu \VEC k \cdot \VEC R_{ji} } J_{ji}^\mp
=   J_{\VEC k}^\mp \quad ,
\end{split}
\end{equation}
and
\begin{equation}\label{eq:Four_J3}
\begin{split}
 J_{\VEC k}^\pm = &
    \sum_i \big(\cos( \VEC k \cdot \VEC R_{ij})+ \iu \sin( \VEC k \cdot \VEC R_{ij}) \big) (J_{ij} \pm \iu D_{ij}^z ) \\ 
 = &\sum_i A_{ij} \cos \left( \VEC k \cdot \VEC R_{ij} \pm \theta_{ij} \right)  \quad , \\ 
\end{split}
\end{equation}
where $\theta_{ij} = \arctan(D_{ij}^z / J_{ij})$ and $A_{ij} = \sqrt{(D_{ij}^z)^2 +(J_{ij})^2}$.
The last equation shows that $J^\pm_\VEC k$ is purely real and, that it can be expanded in terms of cosine functions whose phase is given by the magnetic exchange and DMI ratio.
This derives from the fact that the sum of a Bravais lattice of an antisymmetric function vanishes, $\sum_i \sin(\VEC k\cdot \VEC R_{ij}) J_{ij} = \sum_i \cos(\VEC k \cdot \VEC R_{ij}) D^z_{ij} = 0$.

\subsubsection{Eigenvalues: the frequencies} 
\label{sub:eigen_values_the_frequencies}

\begin{figure}[th!]
\setlength{\unitlength}{1cm} 
\newcommand{\boxsize}{0.3}

  \begin{picture}(9,8)
    
    \put(0.0, 0.){ \includegraphics[width=8.5 cm,trim={0 0 0 0},clip=true]{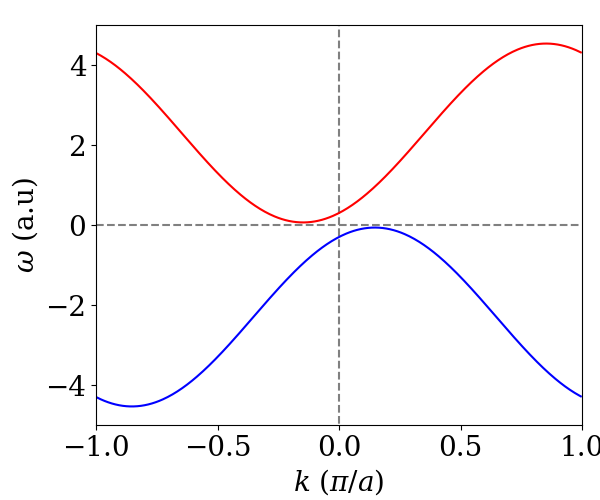} }  
  \end{picture}

  \caption{\label{fig:spin_wave_dispersion_example}
  Spin-wave dispersion for a ferromagnet with DMI.
  The red and blue curves correspond to the functions $\omega^-_{\VEC k}$ and $\omega^+_{\VEC k}$ from Eq.~\eqref{eq:omegapm}, respectively.
  The blue curve is obtained from the red one by time reversal, which enforces $\omega^+_{\VEC k} = -\omega^-_{-\VEC k}$.
  Parameters: $J=1$, $D^z=0.5$, $B=0.3$.
  }
\end{figure}

The differential equations in~\eqref{eq:eq_motion6} have solutions of the type
\begin{equation} \label{eq:time_evolution}
  S_{\VEC k}^\pm(t) = S_{\VEC k}^\pm e^{-\iu \omega^\pm_\VEC k t} \quad ,    
\end{equation}
which plugging into Eqs.~\eqref{eq:eq_motion6} results in
\begin{equation}\label{eq:eq_motion7}
\begin{split}
- \omega^+_\VEC k S_{\VEC k}^+ =& \Big( S  \left( J_{\VEC 0} - J_{\VEC k}^- \right) + B \Big) S_{\VEC k}^+   \\
  \omega^-_\VEC k S_{\VEC k}^- =& \Big( S  \left( J_{\VEC 0} - J_{\VEC k}^+ \right) + B \Big) S_{\VEC k}^- \quad ,  \\
\end{split}
\end{equation}
and therefore, the eigenvalues of these equations that correspond to the solution frequencies are given by
\begin{equation}\label{eq:omegapm}
  \omega^\pm_{\VEC k} =  \mp S  \left( J_{\VEC 0} - J_{\VEC k}^\mp \right) \mp B\quad .
\end{equation}
For a one-dimensional ferromagnet with nearest-neighbour-only MEI and DMI, $J=1$, $D^z=0.5$, and a magnetic field $B=0.3$, we plotted the above equation in Fig.~\ref{fig:spin_wave_dispersion_example}.

Using Eq.~\eqref{eq:Four_J2}, we can notice that
\begin{equation}\label{eq:frequencies_relation}
  \omega^\pm_{-\VEC k}
   = - ( \pm S  \left( J_{\VEC 0} - J_{\VEC k}^\pm \right) \pm B)
   = - \omega^\mp_{\VEC k} \quad ,
\end{equation}
that is, the frequency of each solution is related to the other by an inversion of wavevector and a sign change of the frequency, which can be translated into an inversion of time in Eq.~\eqref{eq:time_evolution}, see Fig.~\ref{fig:spin_wave_dispersion_example}.
Due to $J_{\VEC 0}^\pm = J_\VEC 0$, we have that  $\omega^\pm_{\VEC k}  \rightarrow \mp B$ when $\VEC k \rightarrow 0$.

In the absence of DMI, and if $J_{ij} > 0$, we have that $\omega^\pm_{\VEC k} = \omega^\pm_{-\VEC k} = \pm \omega_\VEC k$, which imply that the frequencies are reciprocally symmetric and additive inverse of each other.
Furthermore, $\omega_\VEC k$ is always real and positive, as we expect for a ferromagnetic system:
\begin{equation}
  \omega_\VEC k = S  \sum_i \left[1 - \cos( \VEC k \cdot \VEC R_{ij} ) \right] J_{ij} + B > 0 \quad .
\end{equation}
For nonzero $D^z_{ij}$, the phases of the cosines change, making the spin-wave dispersion nonreciprocal.

\subsubsection{Local spin dynamics}
\label{sub:local_spin_dynamics}

Now, it is time to transform back, from the circular components to the Cartesian ones in order to understand the precession of individual spins.
For a given wavevector $\VEC k$, we have that
\begin{equation}
\begin{split}
  S_i^+(\VEC k,t) =&\frac{1}{\sqrt{N}} e^{  \iu \VEC k \cdot \VEC R_i}S_{\VEC k}^+(t) 
           =\frac{S_{\VEC k}^+}{\sqrt N} \left( \cos(\VEC k \cdot \VEC R_i - \omega^+_\VEC k t) + \iu \sin(\VEC k \cdot \VEC R_i - \omega^+_\VEC k t) \right) \\
  S_i^-(\VEC k,t) =&\frac{1}{\sqrt{N}} e^{  \iu \VEC k \cdot \VEC R_i}S_{\VEC k}^-(t) 
           =\frac{S_{\VEC k}^-}{\sqrt N} \left( \cos(\VEC k \cdot \VEC R_i - \omega^-_\VEC k t) + \iu \sin(\VEC k \cdot \VEC R_i - \omega^-_\VEC k t) \right) 
           \quad .
\end{split}
\end{equation}
Comparing these equations with their definitions at Eq.~\eqref{eq:circ_components} in terms of the Cartesian components, $ S_i^+ = S_i^x + \iu S_i^y$ and $S_i^- = S_i^x - \iu S_i^y$ , we get a solution for each equation:
\begin{equation}\label{eq:spin_motion}
\begin{split}
  \big( S_i^x , S_i^y \big)(\VEC k) =& \frac{ S_{\VEC k}^+}{\sqrt N} \big( \cos( \VEC k \cdot \VEC R_i - \omega^+_\VEC k t) \hspace{0.1cm}, \hspace{0.1cm} \sin(\VEC k \cdot \VEC R_i - \omega^+_\VEC k t) \big) \\
  \big( S_i^x , S_i^y \big)(\VEC k) =& \frac{ S_{\VEC k}^-}{\sqrt N} \big( \cos(-\VEC k \cdot \VEC R_i + \omega^-_\VEC k t) \hspace{0.1cm}, \hspace{0.1cm} \sin(-\VEC k \cdot \VEC R_i + \omega^-_\VEC k t) \big) \\
\end{split}
\end{equation}
We can show that these two equations are equivalent by substituting one of the amplitude by its dual: $\left( S_{\VEC k}^\pm \right)^* = S_{-\VEC k}^\mp $, see Eq.~\eqref{eq:dual_Sk}, and using the relation derived in Eq.~\eqref{eq:frequencies_relation}.
Doing so in the second solution of Eq.~\eqref{eq:spin_motion}, we get
\begin{equation}
\begin{split}
  \big( S_i^x , S_i^y \big)(-\VEC k) =& \frac{ (S_{\VEC k}^+)^* }{\sqrt N} \big( \cos(\VEC k \cdot \VEC R_i - \omega^+_\VEC k t) \hspace{0.1cm} , \hspace{0.1cm} \sin(\VEC k \cdot \VEC R_i - \omega^+_\VEC k t) \big) \quad ,
\end{split}
\end{equation}
which is, considering $S_{\VEC k}^+$ real, equivalent to the first solution in Eq.~\eqref{eq:spin_motion} but with opposite wavevector.
These solutions represent counterclockwise circular precessions, and they are related to each other by a time inversion!


\subsection{Circular components duality} 
\label{sec:ciarcular_components_duality}

The following reviews the duality between the circular components of the spin moments and how they evolve through the transformation considered previously, such as the Fourier transformation. 

From the definition of the circular components in Eq.~\eqref{eq:circ_components}, we have that
\begin{equation}
  \left( S_i^\pm \right)^* = S_i^\mp \quad, 
\end{equation}
that is, one is the complex conjugate of the other.
Given the Fourier transformation definitions by Eq.~\eqref{eq:Four_transf}, the complex conjugate duality of the Fourier counterparts is given by
\begin{equation}\label{eq:dual_Sk}
  \left( S_{\VEC k}^\pm \right)^* = S_{-\VEC k}^\mp \quad.
\end{equation}
Given the definition of the time evolution, Eq.~\eqref{eq:time_evolution}, we have that 
\begin{equation}
\left( S_{\VEC k}^\pm(t) \right)^*  = S_{-\VEC k}^\mp (-t)  \quad .
\end{equation}
And again, we see that they are related by a time reversal operation.


\end{appendix}

\end{document}